\PassOptionsToPackage{hyperfootnotes=false}{hyperref}
\documentclass[11pt]{article}

\usepackage{authblk}
\usepackage{orcidlink}

\usepackage[T1]{fontenc}
\usepackage{lmodern}
\usepackage[margin=1.07in]{geometry}
\usepackage{microtype}
\usepackage{amsmath,amssymb,amsthm,mathtools}
\usepackage{booktabs,tabularx,array}
\usepackage{enumitem}
\usepackage{tikz}
\usetikzlibrary{arrows.meta,positioning,calc,fit}
\usepackage[numbers,sort&compress]{natbib}
\usepackage[dvipsnames]{xcolor}
\definecolor{myRed}{rgb}{0.82,0.13,0.56}
\definecolor{myBlue}{RGB}{13,55,174}
\usepackage{hyperref}
\hypersetup{
  colorlinks=true,
  citecolor=myRed,
  linkcolor=myBlue,
  urlcolor=PineGreen
}
\usepackage[noabbrev,capitalise]{cleveref}
\definecolor{ink}{HTML}{17334B}
\definecolor{tealfig}{HTML}{087E81}
\definecolor{bluefig}{HTML}{34649A}
\definecolor{ochre}{HTML}{AE5A18}
\definecolor{muted}{HTML}{536777}
\tikzset{flow/.style={-{Latex[length=2mm]},line width=.65pt,draw=muted},
  box/.style={draw=ink!50,fill=ink!4,rounded corners=2pt,align=center,inner sep=6pt},
  point/.style={circle,fill=ink,inner sep=1.7pt},
  figuretext/.style={font=\small,text=ink}}
\allowdisplaybreaks[2]
\setlist{itemsep=2pt,topsep=4pt}

\newtheorem{theorem}{Theorem}[section]
\newtheorem{lemma}[theorem]{Lemma}
\newtheorem{proposition}[theorem]{Proposition}
\newtheorem{corollary}[theorem]{Corollary}
\newtheorem{claim}[theorem]{Claim}
\newtheorem{fact}[theorem]{Fact}
\theoremstyle{definition}
\newtheorem{definition}[theorem]{Definition}
\newtheorem{example}[theorem]{Example}

\crefname{equation}{equation}{equations}
\crefname{figure}{Figure}{Figures}
\crefname{table}{Table}{Tables}
\crefname{section}{Section}{Sections}
\crefname{appendix}{Appendix}{Appendices}
\crefname{fact}{Fact}{Facts}
\crefname{claim}{Claim}{Claims}

\newcommand{\Z}{\mathbb Z}
\newcommand{\N}{\mathbb N}
\newcommand{\PSPACE}{\mathsf{PSPACE}}

\newcommand{\NP}{\mathsf{NP}}
\newcommand{\coNP}{\mathsf{coNP}}
\newcommand{\DTIME}{\mathsf{DTIME}}
\DeclareMathOperator{\lcm}{lcm}
\newcommand{\ap}[2]{[#1]_{#2}}
\newcommand{\fp}{\mathrm{Fp}}
\newcommand{\lam}{\Lambda}

\newcommand{\pv}{\varpi}

\title{Pinwheel Scheduling is PSPACE-Complete}
\author[1,2]{Dimitris Fotakis \orcidlink{0000-0001-6864-8960} }
\author[1,2]{Sotiris Kanellopoulos \orcidlink{0009-0006-2999-0580} }
\author[3]{Giorgos Mitropoulos \orcidlink{0009-0000-0479-263X} }
\author[1,2]{\\ Aris Pagourtzis \orcidlink{0000-0002-6220-3722} }
\author[2,4]{Christos Papadimitriou \orcidlink{0009-0000-7264-8015} }
\author[1,2]{Christos Pergaminelis \orcidlink{0009-0009-8981-3676} }
\author[1,2]{\\ Thanos Tolias \orcidlink{0009-0003-8354-8855} }

\affil[1]{National Technical University of Athens, 15780 Athens, Greece}
\affil[2]{Archimedes, Athena RC, 15125 Marousi, Greece}
\affil[3]{Sorbonne Université, CNRS, LIP6, F-75005 Paris, France}
\affil[4]{Columbia University, 2960 Broadway, New York, NY 10027, USA}

\affil[ ]{\texttt{
\href{mailto:fotakis@cs.ntua.gr}{fotakis@cs.ntua.gr} \quad
\href{mailto:s.kanellopoulos@athenarc.gr}{s.kanellopoulos@athenarc.gr} \quad
\href{mailto:georgios.mitropoulos@lip6.fr}{georgios.mitropoulos@lip6.fr} \quad
\href{mailto:pagour@cs.ntua.gr}{pagour@cs.ntua.gr} \quad
\href{mailto:christos@columbia.edu}{christos@columbia.edu} \quad
\href{mailto:chr.pergaminelis@athenarc.gr}{chr.pergaminelis@athenarc.gr} \quad
\href{mailto:thanostolias@mail.ntua.gr}{thanostolias@mail.ntua.gr}
}}

\date{}
\hypersetup{pdftitle={The PSPACE-Completeness of Pinwheel Scheduling},pdfauthor={}}

\begin{document}
\maketitle

\renewcommand\thefootnote{}\footnotetext{Dimitris Fotakis and Thanos Tolias were supported by the ERC Consolidator Grant COMFORMAL (No. 101232091). Sotiris Kanellopoulos and Christos Pergaminelis acknowledge the financial support of the Archimedes Research Unit (Athena RC) during their PhD.}
\setcounter{footnote}{0}\renewcommand\thefootnote{\arabic{footnote}}

\begin{abstract}
Pinwheel Scheduling is a classical periodic scheduling problem, receiving as input the deadlines $d_i$ of $n$ tasks and asking for an infinite schedule with one task per time unit, such that every task $i$ is scheduled at least once in every $d_i$ time units. Holte, Mok, Rosier, Tulchinsky, and Varvel introduced the problem in 1989 and showed that it lies in \(\PSPACE\). Its exact complexity has been open since then. \(\NP\)-hardness for explicitly listed deadlines was established only recently, and membership in \(\NP\) is open.

In this work we prove that Pinwheel Scheduling is \(\PSPACE\)-complete when the deadlines are written in binary. This finally settles the long-standing open question about the complexity of Pinwheel Scheduling. 
We present a chain of reductions, starting from Periodic 3-SAT, a classical \(\PSPACE\)-complete problem, and using two newly-defined intermediate periodic scheduling problems. 
These intermediate problems can be seen as periodically assigning tasks to days, with each day containing a fixed number of slots (\emph{hours}) in which tasks are scheduled. In the first of these problems, each day is additionally allowed to contain \emph{breaks}, that is, hours in which no task can be scheduled.

\end{abstract}

\setcounter{tocdepth}{2}
\newpage

\section{Introduction}\label{sec:intro}

A recurring task can be easy to perform and still be hard to schedule. In Pinwheel Scheduling, one machine performs unit-time jobs, and job \(i\) may go unserved for at most \(d_i\) consecutive time slots. The task can be the reception of a satellite transmission at a ground station, the broadcast of a data page, a maintenance operation, or a visit to a location with a fixed turnover time \citep{holte1989,barnoyladner2003,bosman2022,kawamura2025perpetual}. There are no release dates, durations, or initial phases in the input. The only data are the integers \(d_1,\ldots,d_n\), and we ask whether one job per slot can be chosen forever so that every job occurs in every window of \(d_i\) consecutive slots.

The \emph{density} \(\rho=\sum_i 1/d_i\) is the fraction of slots that the jobs require at minimum, so \(\rho\le1\) is necessary. Kawamura proved that \(\rho\le 5/6\) is sufficient \citep{kawamura2024,kawamura2026pnas}, which settled the conjecture of Chan and Chin \citep{chan1993}. The bound is sharp: the instance \((2,3,d)\) has density \(5/6+1/d\) and is infeasible for every \(d\). Indeed, an occurrence of the third job forces both of its neighbours to be the deadline-two job, and then the three slots around it contain no occurrence of the deadline-three job. Between \(5/6\) and \(1\), feasibility depends on the individual deadlines, and the question is how hard it is to decide.

\paragraph{Complexity status.}
Holte, Mok, Rosier, Tulchinsky, and Varvel introduced the problem and showed that it belongs to \(\PSPACE\) \citep{holte1989}. A feasible instance has a periodic schedule, and feasibility amounts to a cycle in an exponentially large graph of job ages. For more than three decades this remained the only general upper bound, and no lower bound was known for the standard encoding, in which the jobs are listed explicitly and the deadlines are written in binary. Holte et al.\ claimed \(\NP\)-hardness when equal deadlines are grouped and their multiplicities are written in binary. The proof published by Bar-Noy, Ladner, and Tamir \citep{barnoy2007} contains an error found by Jacobs and Longo \citep{jacobs2014}, and a correct proof was given by Kleinberg and Mishra \citep{kleinberg2026}. For the explicit encoding, Jacobs and Longo ruled out pseudopolynomial algorithms unless SAT has randomized algorithms running in time \(n^{O(\log n\log\log n)}\) \citep{jacobs2014}. Kobayashi and Lin ruled out polynomial algorithms for instances of density one unless \(\NP\subseteq\DTIME(n^{O(\log n)})\) \citep{kobayashi2025}, and the journal version with Swernofsky improves the assumption to \(\NP\subseteq\DTIME(n^{O(\log\log n)})\) \citep{kobayashi2026tcs}. In 2026, Kleinberg and Mishra proved that Pinwheel Scheduling is \(\NP\)-hard, even at density one \citep{kleinberg2026}, and Kobayashi, Lin, and Swernofsky proved that it is strongly \(\NP\)-complete at density one \citep{kobayashi2026dense}. Whether the general problem is in \(\NP\), and whether it is \(\PSPACE\)-hard, have remained open \citep{kanellopoulos2026soda,kanellopoulos2026icalp,kanellopoulos2026cover}.

\paragraph{Our result.}
We settle the question.

\begin{theorem}\label{thm:main}
Pinwheel Scheduling is \(\PSPACE\)-complete under polynomial-time many-one reductions, for one machine, unit-time jobs, an explicit list of jobs, and deadlines written in binary.
\end{theorem}

The reduction outputs polynomially many jobs, and their deadlines have polynomially many bits. Some deadlines are exponentially large in value, so our hardness is weak in the sense of strong versus weak \(\NP\)-hardness
. Since \(\NP\) and \(\coNP\) are contained in \(\PSPACE\), we obtain the following corollary.

\begin{corollary}\label{cor:intro-consequences}
Pinwheel Scheduling with explicit binary input does not belong to \(\NP\) unless \(\NP=\PSPACE\) (resp. to \(\coNP\) unless \(\coNP=\PSPACE\)).
\end{corollary}

The hardness concentrates just below density one.

\begin{theorem}\label{thm:intro-near}
For every fixed rational \(\epsilon>0\), Pinwheel Scheduling with explicit binary input remains \(\PSPACE\)-complete on instances with \(1-\epsilon<\rho<1\).
\end{theorem}

At density exactly one, every job of a feasible instance must recur with gap exactly \(d_i\) in a periodic schedule, so one phase per job is a polynomial certificate \citep{holte1989}. \Cref{thm:intro-near} therefore identifies the transition from \(\NP\)- to \(\PSPACE\)-completeness precisely: any positive  slack, however small, suffices.

\subsection{A note on below-one density}\label{sec:why}

At density one, the phases of the jobs are static. Below density one they can move. Our proof turns this motion into a memory. To make this motion usable, we group consecutive slots into \emph{days} of size $w$, numbered $1,\ldots,w$ within each day and called \emph{hours}. One can think of this structure as a matrix with infinitely many rows and $w$ columns, where rows represent days and columns represent hours. A job has an exact period \(b_i\) measured in days, so it occurs in exactly one slot of every \(b_i\)-th day. Its hour, however, may change from visit to visit, subject to an \emph{ascent bound} \(\delta_i\): the hour may rise by at most \(\delta_i\) between consecutive visits, and it may fall arbitrarily. We call this model \emph{Daily Scheduling}. In a pinwheel instance, a deadline of the form \(d_i=Tb_i+\delta_i\), with days of length \(T\), allows precisely this behaviour: a job at hour \(y\) of day \(t\) and hour \(y'\) of day \(t+b_i\) has gap \(Tb_i+y'-y\), which is at most \(d_i\) exactly when \(y'-y\le\delta_i\).


\subsection{Technical overview}\label{sec:overview}

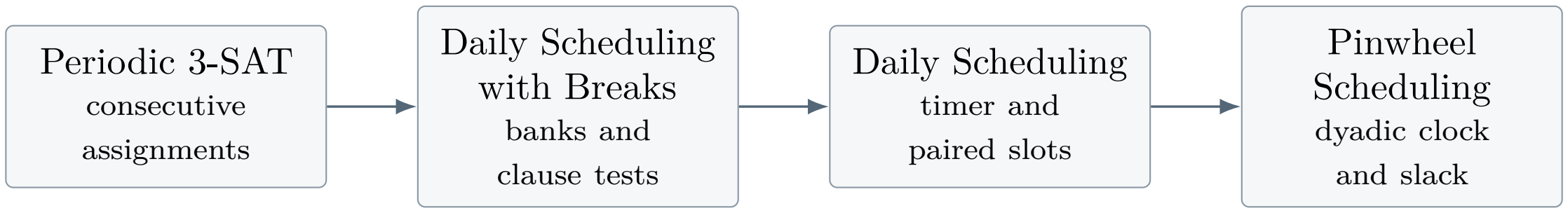
\begin{figure}[t]
\centering
\begin{tikzpicture}[node distance=8.5mm, every node/.style={font=\small}]
  \node[box,minimum height=15mm,text width=2.6cm] (sat) {Periodic 3-SAT\\[-1pt]\scriptsize consecutive assignments};
  \node[box,right=of sat,minimum height=15mm,text width=2.6cm] (mask) {Daily Scheduling\\with Breaks\\[-1pt]\scriptsize banks and clause tests};
  \node[box,right=of mask,minimum height=15mm,text width=2.6cm] (cal) {Daily Scheduling\\[-1pt]\scriptsize timer and paired slots};
  \node[box,right=of cal,minimum height=15mm,text width=2.6cm] (pin) {Pinwheel Scheduling\\[-1pt]\scriptsize dyadic clock and slack};
  \draw[flow] (sat) -- (mask);
  \draw[flow] (mask) -- (cal);
  \draw[flow] (cal) -- (pin);
\end{tikzpicture}
\caption{The three reductions. The two intermediate models use ordered hours. Only the first has prescribed breaks.}\label{fig:reductions}
\end{figure}

The proof consists of three polynomial-time reductions, shown in \cref{fig:reductions}. Each preserves feasibility in both directions.

\paragraph{Step 1: Periodic 3-SAT to Daily Scheduling with Breaks (\cref{sec:masked}).}
We reduce from Periodic 3-SAT. Given a 3-CNF formula \(F(X,Y)\) over two copies of a variable set, it asks for an infinite sequence of assignments \(X_0,X_1,\ldots\) with \(F(X_t,X_{t+1})\) true for every \(t\). The problem is \(\PSPACE\)-complete by a theorem of Orlin \citep{orlin1981}. It suits periodic scheduling because, as in a pinwheel instance, no initial state is prescribed. In Daily Scheduling with Breaks, periodic \emph{breaks} determine which hours are available each day. Two banks of jobs store two consecutive assignments, one strip of hours per variable. A strip holds one job of ascent bound one and one of ascent bound two, and the variable is true when the bound-two job is on top. A clause is checked in four days. A checker job must descend from a guard band, pass through the strip of one of the clause's literals, and return to a guard band; two companion jobs and an exact counting argument force it to stop exactly at a port whose literal is true. Writes are produced by a waveform of available hours that first verifies that each strip holds one job of each kind and then lets either job end on top. One period of the break calendar reads \(F(A,B)\), rewrites \(A\), reads \(F(B,A)\), and rewrites \(B\).

\paragraph{Step 2: removing the breaks (\cref{sec:compiler}).}
Pinwheel instances have no prescribed breaks, so we enforce them using exact periods. Every physical hour receives a private prime \(\pi_\ell\) and \(\pi_\ell-1\) rigid jobs of period \(\pi_\ell\), which pin a distinct label to each hour and leave one residue class free. Nested \emph{selector splits} carve the free class into reservations, each identified by a private prime power. Five jobs whose periods are indexed by the vertices of a five-cycle test that two reservations share a time class, and, with ascent bound one, that their hours are adjacent. These tests force the physical order of the labels up to a global reflection, and they copy a row residue modulo \(M\) to every hour. A moving timer job with ascent bound one must climb the timer hours one step at a time, so it forces the rows into their cyclic order. Two difficulties remain. First, refining reservations to a common modulus does not preserve the number of vacancies at each hour, because refiners may migrate. We give each logical slot two physical copies and a private prime \(q_j\), and we show, using only global saturation, that the pair has exactly one vacancy per period. Second, a moving job may wander into foreign reservations; we only need it to behave on a long interval, and we find such \emph{clean windows} by translating by multiples of a period that is coprime to all foreign prime powers. On a clean window, the computation jobs project onto a schedule of the Daily Scheduling with Breaks instance.

\paragraph{Step 3: from Daily Scheduling to Pinwheel Scheduling (\cref{sec:dyadic}).}
For each level \(h=1,\ldots,k\) we add \(w\) clock jobs with deadline \(2^hw\). Deleting the clock jobs level by level shows that any window of \(T=2^kw\) slots contains at most \(w\) other occurrences. Each job of the Daily Scheduling instance gets the deadline \(Tb_i+\delta_i\). The output density is strictly below one, but only by \(O(n\Delta/T^2)\), where \(\Delta=\max_i\delta_i\). An averaging argument finds a long window whose total deviation from exact days is small, a common cut turns that window into exact days, and a finite-state repetition yields an infinite schedule for Daily Scheduling.

\subsection{Related work}\label{sec:related}

\paragraph{Density thresholds and algorithms.}
Holte et al.\ showed that density at most \(1/2\) suffices and that instances with two distinct deadlines are easy \citep{holte1989,holte1992}. Chan and Chin developed schedulers based on period reductions and conjectured the threshold \(5/6\) \citep{chan1992,chan1993}. Fishburn and Lagarias studied achievable densities \citep{fishburn2002}. G\k{a}sieniec, Smith, and Wild verified the \(5/6\) conjecture for instances with at most \(12\) jobs and showed that the problem is fixed-parameter tractable in the number of jobs \citep{gasieniec2022}. Kawamura proved the conjecture \citep{kawamura2024,kawamura2026pnas}. These results concern sufficient conditions and the construction of schedules. They leave a density range in which feasibility depends on the individual deadlines.

\paragraph{Exact periods and density one.}
In the \emph{periodic maintenance} problem, each job must recur with gap exactly \(d_i\). Two exact jobs with periods \(a,b\) and phases \(s,t\) collide exactly when \(s\equiv t\pmod{\gcd(a,b)}\), so one phase per job is a polynomial certificate. The problem is strongly \(\NP\)-complete \citep{mok1989,korst1996,barnoy2002}. At density one, every feasible pinwheel schedule is exact \citep{holte1989}, which places dense pinwheel scheduling in \(\NP\). The recent hardness results all work in this regime. Kleinberg and Mishra pad an exact-period reduction from a bounded-occurrence version of 3-SAT to density one with polynomially many jobs of exponential value, and thereby prove weak \(\NP\)-hardness \citep{kleinberg2026}; they also give an efficient approximation scheme for a gap version of the problem, which yields one for Bamboo Garden Trimming. Kobayashi, Lin, and Swernofsky prove strong \(\NP\)-completeness at density one \citep{kobayashi2026dense}. Our instances have density strictly below one, as they must, and their phases change over time.

\paragraph{Finite and varying versions.}
Kanellopoulos, Pergaminelis, Kokkou, Markou, and Pagourtzis introduced \(k\)-Visits, in which each job is executed exactly \(k\) times, and proved strong \(\NP\)-completeness for \(k=2\) \citep{kanellopoulos2026soda}. As a corollary, Pinwheel Scheduling becomes strongly \(\NP\)-hard when each deadline may change once after a given number of executions. Kanellopoulos, Mitropoulos, Pergaminelis, and Tolias extended the hardness to maximum multiplicity two and studied density thresholds and tractable cases \citep{kanellopoulos2026icalp}, and Kanellopoulos introduced the covering analogue \citep{kanellopoulos2026cover}. The finite versions check a single bounded schedule. Our reduction must check an unbounded sequence of assignments with one fixed list of jobs.

\paragraph{\(\PSPACE\)-hardness of recurrent problems.}
Orlin proved that periodic versions of \(\NP\)-complete problems, in particular Periodic SAT, are \(\PSPACE\)-complete \citep{orlin1981}. For real-time task systems, Baruah, Howell, and Rosier showed that feasibility on one processor is \(\coNP\)-complete in the strong sense, with higher complexity when start times are not given \citep{baruah1993}. Closest to us, Ho and Ouaknine proved that cyclic routing of a single UAV, with revisit deadlines and a matrix of travel times between targets, is \(\PSPACE\)-complete \citep{ho2015,drucker2019}. They also reduce from Periodic SAT, using two traversal directions per variable and consistency gadgets between iterations. Their instances depend on the input travel times. In Pinwheel Scheduling every slot has the same length and the input consists of the deadlines alone. The calendar compiler of \cref{sec:compiler} is what generates, from deadlines alone, the restrictions that the bank and checker jobs need.

\paragraph{Covering and related models.}
Pinwheel Covering asks that job \(i\) occur at most once in every \(d_i\) consecutive slots while every slot is used \citep{kawamura2020,kawamura2025cover}. At density one it coincides with the packing version \citep{kobayashi2025}. Its optimal density bound was determined independently by Kawamura and Kobayashi \citep{kawamura2026cover} and by Mishra \citep{mishra2026}. Windows scheduling generalizes Pinwheel Scheduling to several machines \citep{barnoyladner2003,barnoy2007}. Jacobs and Longo proved \(\coNP\)-hardness for unit jobs on several machines with migration \citep{jacobs2014}. Our spacing and rounding arguments use maximum gaps, and we do not know whether they transfer to covering.

\Cref{tbl:complexity} summarizes the known complexity results by model and encoding.

\begin{table}[t]
\centering
\small
\begin{tabularx}{\linewidth}{@{}p{.30\linewidth}p{.27\linewidth}X@{}}
\toprule
Model & Encoding or restriction & Complexity \\
\midrule
Exact periods, one machine & Explicit list & Strongly \(\NP\)-complete \citep{mok1989,korst1996,barnoy2002} \\
Pinwheel, \(\rho=1\) & Explicit list, binary & \(\NP\)-complete \citep{kleinberg2026} \\
Pinwheel, \(\rho=1\) & Explicit list, unary & \(\NP\)-complete \citep{kobayashi2026dense} \\
Finite \(2\)-Visits & Explicit list, unary & \(\NP\)-complete \citep{kanellopoulos2026soda,kanellopoulos2026icalp} \\
Pinwheel, one deadline change & Explicit list, unary & \(\NP\)-hard \citep{kanellopoulos2026soda} \\
Pinwheel, general & Binary multiplicities & \(\NP\)-hard \citep{kleinberg2026}; \newline\(\PSPACE\)-hard (\cref{cor:compact}) \\
Pinwheel, general & Explicit list, binary & \(\NP\)-hard \citep{kleinberg2026};\newline\(\PSPACE\)-complete (\cref{thm:main}) \\
Pinwheel, \(1-\epsilon<\rho<1\) & Explicit list, binary & \(\PSPACE\)-complete (\cref{thm:intro-near}) \\
Pinwheel, general & Explicit list, unary & In \(\PSPACE\); strongly \(\NP\)-hard \citep{kobayashi2026dense};\newline\(\PSPACE\)-hardness open \\
\bottomrule
\end{tabularx}
\caption{Complexity by recurrence rule and input encoding. An explicit list counts repeated deadlines as separate jobs. Unary hardness is strong hardness for the binary encoding.}\label{tbl:complexity}
\end{table}

\subsection{Organization}

\Cref{sec:prelim} defines Pinwheel Scheduling, Daily Scheduling, and Daily Scheduling with Breaks and proves their basic properties. \Cref{sec:source} states the source problem. \Cref{sec:masked,sec:compiler,sec:dyadic} contain the three reductions. \Cref{sec:consequences} proves the main theorem and its corollaries.
Appendix~\ref{app:notation} offers a summary of the notation.

\section{Preliminaries}\label{sec:prelim}


For a positive integer \(m\) and an integer \(a\), we write \(\ap{a}{m}=\{t\in\Z:t\equiv a\pmod m\}\) and call it a \emph{class modulo \(m\)}. We write \([k]=\{1,\ldots,k\}\). Intervals of integers are written \(\{a,\ldots,b\}\). We use the Chinese remainder theorem (CRT) in the following form.

\begin{fact}\label{fact:crt}
For positive integers \(p,q\) and integers \(a,b\), the set \(\ap ap\cap\ap bq\) is nonempty if and only if \(a\equiv b\pmod{\gcd(p,q)}\). In that case it is a single class modulo \(\lcm(p,q)\).
\end{fact}

We often use the following consequence: if \(Z\) is a class modulo \(m\) and \(Q\) is a positive integer, then \(Z\) meets exactly \(Q/\gcd(m,Q)\) classes modulo \(Q\). In particular, \(Z\) lies inside a single class modulo \(Q\) if and only if \(Q\mid m\).

Input job identities are distinct even when two jobs have the same parameters. All schedules in this paper are two-sided, that is, indexed by \(\Z\). By \cref{prop:space,lem:frames} below, one-sided schedules indexed by \(\N\) lead to the same feasibility questions.

\subsection{Pinwheel Scheduling}

An instance of \emph{Pinwheel Scheduling} is a list of positive integers \(d_1,\ldots,d_n\), the \emph{deadlines}. A \emph{schedule} is a map \(S:\Z\to[n]\cup\{\bot\}\), where \(\bot\) marks an idle slot. It is \emph{feasible} if every window \(\{s,\ldots,s+d_i-1\}\) contains a slot \(s'\) with \(S(s')=i\), for every job \(i\) and every \(s\in\Z\). Equivalently, every job occurs infinitely often in both directions and every gap between consecutive occurrences of job \(i\) is at most \(d_i\). An idle slot can always be given to an arbitrary job without destroying feasibility. The \emph{density} of the instance is \(\rho=\sum_i1/d_i\). In the decision problem, the deadlines are listed explicitly, one entry per job, and written in binary.

\begin{proposition}[\citealp{holte1989}]\label{prop:space}
Pinwheel Scheduling belongs to \(\PSPACE\). Every feasible instance has a periodic feasible schedule.
\end{proposition}


\subsection{Daily Scheduling}

\begin{definition}[Daily Scheduling]\label{def:ref}
An instance consists of a width \(w\ge1\) and jobs \(i=1,\ldots,n\), each with an integer \emph{period} \(b_i\ge1\) and an integer \emph{ascent bound} \(\delta_i\ge0\). Time is divided into \emph{days} indexed by \(\Z\), and each day has the \emph{hours} \(1,\ldots,w\). A schedule chooses a \emph{phase} \(\alpha_i\bmod b_i\) for every job and, for every day \(t\in\ap{\alpha_i}{b_i}\), an hour \(y_i(t)\in[w]\). We say that job \(i\) is \emph{due} on the days of \(\ap{\alpha_i}{b_i}\). The schedule is feasible if
\begin{enumerate}[label=(\roman*)]
\item jobs due on the same day receive distinct hours, and
\item \(y_i(t+b_i)-y_i(t)\le\delta_i\) for every job \(i\) and every \(t\in\ap{\alpha_i}{b_i}\).
\end{enumerate}
There is no bound on downward moves. The instance is \emph{saturated} if \(\sum_i1/b_i=w\). A job with \(\delta_i=0\) is \emph{rigid}, and all other jobs are \emph{moving}. A \emph{slot} is a pair (hour, day).
\end{definition}

\begin{definition}[Daily Scheduling with Breaks]\label{def:masked}
An instance of width \(w\) additionally specifies a \emph{calendar period} \(M\ge1\) and an explicit list of sets \(S_0,\ldots,S_{M-1}\subseteq[w]\), the \emph{rows} of the calendar. Each row \(S_t\) is given by a \(w\)-bit vector whose \(x\)-th bit is \(1\) exactly when \(x\in S_t\). On days congruent to \(t\) modulo \(M\), the hours in \(S_t\) are available and the hours in \([w]\setminus S_t\) are \emph{breaks}. Every period satisfies \(b_i\mid M\). In a feasible schedule, a job due on day \(t\) must also use an hour of \(S_{t\bmod M}\). The instance is \emph{saturated} if \(\sum_i1/b_i=\frac1M\sum_{t=0}^{M-1}|S_t|\).
\end{definition}

The first model has \emph{no breaks}. The input size of a Daily Scheduling with Breaks instance includes the explicit calendar, so the reductions below may run in time polynomial in \(M\) and \(w\).

\begin{lemma}
\label{lem:frames}
Fix the phases of all jobs. Let \(L=\lcm_ib_i\) in Daily Scheduling, and let \(L=M\) in Daily Scheduling with Breaks. The \emph{state} before day \(t\) is the residue \(t\bmod L\) together with the hour of every job at its last visit before \(t\). Let \(b_*=\max_ib_i\).
\begin{enumerate}[label=(\alph*)]
\item If a schedule with these phases satisfies conditions (i) and (ii) and respects any prescribed breaks on more than \(b_*+Lw^n\) consecutive days, then the instance has a periodic feasible schedule with the same phases.
\item Every feasible instance has a periodic feasible schedule.
\item In a feasible saturated instance, every available hour of every day is occupied, in every feasible schedule.
\end{enumerate}
\end{lemma}
\begin{proof}
(a) After the first \(b_*\) days of the interval every job has been due once, so the state is defined. There are at most \(Lw^n\) states, so two days \(t_1<t_2\) of the interval have the same state. Then \(t_1\equiv t_2\pmod L\), and every period \(b_i\) divides \(t_2-t_1\). Repeat the days \(\{t_1,\ldots,t_2-1\}\) with period \(t_2-t_1\) in both directions. Inside a copy, every constraint holds because it held in the original schedule. Consider a job \(i\) and its last visit before \(t_2\). It occurs on day \(t_2-e\) for some \(e\in[b_i]\), since \(t_2\equiv t_1\pmod{b_i}\), and its hour equals the hour of the last visit before \(t_1\), by the equality of states. The next visit in the repeated schedule is the copy of the first visit at or after \(t_1\). In the original schedule, that visit follows the last visit before \(t_1\) with a legal ascent and exactly \(b_i\) days later. Hence the constraint across the boundary holds. The breaks repeat since \(M\mid t_2-t_1\) in Daily Scheduling with Breaks.

(b) A feasible schedule satisfies the hypothesis of (a) on every interval.

(c) Let \(N(t)\) be the number of jobs due on day \(t\), and let \(R(t)\) be the number of available hours, namely \(w\) or \(|S_{t\bmod M}|\). Both functions are periodic with period \(L\), and \(N(t)\le R(t)\) by condition (i). Over one period, \(N\) has average \(\sum_i1/b_i\) and \(R\) has the same average by saturation. A nonnegative integer function \(R-N\) with average zero vanishes everywhere.
\end{proof}

\begin{lemma}\label{lem:rigid}
In a periodic feasible schedule, every rigid job \(j\) stays at one hour \(\rho_j\). Its \emph{footprint} is the set of slots \(\fp(j)=\{\rho_j\}\times\ap{\alpha_j}{b_j}\). Footprints of distinct jobs are disjoint. Two rigid jobs with periods \(p,p'\) at the same hour have disjoint footprints if and only if their phases differ modulo \(\gcd(p,p')\).
\end{lemma}
\begin{proof}
The hour of a rigid job does not increase from one visit to the next. In a periodic schedule it returns to its initial value, so it is constant. The last claim is \cref{fact:crt}.
\end{proof}

All later arguments about rigid jobs use a periodic witness, which exists by \cref{lem:frames}.

\subsection{Periodic 3-SAT}\label{sec:source}

Let \(X=(x_1,\ldots,x_N)\) and \(Y=(y_1,\ldots,y_N)\) be two copies of a set of Boolean variables. An instance of \emph{Periodic SAT} is a CNF formula \(F(X,Y)\). It is satisfiable if there is an infinite sequence of assignments \(X_0,X_1,\ldots\in\{0,1\}^N\) such that \(F(X_t,X_{t+1})\) holds for every \(t\ge0\). Equivalently, \(F\) is satisfiable if and only if the directed graph \(\mathcal G_F\) on \(\{0,1\}^N\) with an edge \(x\to y\) whenever \(F(x,y)\) holds contains a directed cycle: an infinite walk in a finite graph repeats a vertex, and a cycle can be traversed forever. In particular, a satisfiable formula has a periodic, two-sided satisfying sequence \((X_t)_{t\in\Z}\). We stress that no initial assignment is prescribed. This matches Pinwheel Scheduling, where no initial state of the schedule is prescribed either.

\begin{theorem}[\citealp{orlin1981}]\label{thm:orlin}
Periodic SAT is \(\PSPACE\)-complete.
\end{theorem}

We use the following normal form. We call \(x_k\) and \(y_k\) different variables. \emph{Periodic 3-SAT} is the restriction of Periodic SAT to formulas in which every clause has exactly three literals on three different variables.

\begin{lemma}[\citealp{orlin1981}]\label{lem:psat}
Periodic 3-SAT is \(\PSPACE\)-complete.
\end{lemma}

From now on, \(F\) is a Periodic 3-SAT formula with \(N\) variables in each copy and \(m\) clauses \(C_1,\ldots,C_m\).

\section{From Periodic 3-SAT to Daily Scheduling with Breaks}\label{sec:masked}

\begin{theorem}\label{thm:masked}
Given a Periodic 3-SAT formula \(F\) with \(N\) variables per copy and \(m\) clauses, one can compute in polynomial time a saturated instance of Daily Scheduling with Breaks satisfying the following properties. All periods are \(1\) or \(2\); the width is \(w=72N+15\); the calendar period is \(M=4(2m+8)\); every row has exactly \(4N+3\) available hours; and the instance is feasible if and only if \(F\) is satisfiable.
\end{theorem}

\Cref{sec:masked-construction} gives the construction, \cref{sec:masked-complete} proves completeness, and \cref{sec:masked-sound} proves soundness.

\subsection{Construction}\label{sec:masked-construction}

\paragraph{Geometry.}
There are two \emph{banks}, \(\mathsf A\) and \(\mathsf B\). Each bank has one \emph{strip} per variable index \(k\in[N]\), so there are \(v=2N\) strips. We index them by \(\sigma\in\{0,\ldots,v-1\}\): strip \(\sigma=k-1\) belongs to variable \(k\) of bank \(\mathsf A\), and strip \(\sigma=N+k-1\) to variable \(k\) of bank \(\mathsf B\). Let
\[
W=12v+2,\qquad G=W+4,\qquad o=G+1,\qquad o_\sigma=o+12\sigma .
\]
Strip \(\sigma\) owns the hours \(R_\sigma=\{o_\sigma,\ldots,o_\sigma+8\}\). Its \emph{low hour} is \(o_\sigma\) and its \emph{high hour} is \(o_\sigma+6\); together they are its \emph{ordinary hours}. The three \emph{parking hours} are \(p_1=o+12v\), \(p_2=p_1+1\), and \(p_3=p_1+2\). The strips and the parking hours form the \emph{middle band} \(\{o,\ldots,o+W\}\). For an hour \(u\) of the middle band, \(u-G\) is its \emph{lower guard} and \(u+G\) its \emph{upper guard}. The lower guards lie in \(\{1,\ldots,W+1\}\) and the upper guards in \(\{o+G,\ldots,o+W+G\}\). The width is \(w=o+W+G=3W+9=72N+15\).

We call the strips \(R_\sigma\), the parking set \(\{p_1,p_2,p_3\}\), the lower guard band, and the upper guard band the \emph{regions}. If one region lies above another, then every hour of the upper region exceeds every hour of the lower region by at least \(4\). For example, \(o_{\sigma+1}-(o_\sigma+8)=4\), \(p_1-(o_{v-1}+8)=4\), \(o-(o+W-G)=4\), and \((o+G)-(o+W)=4\).

\paragraph{Jobs.}
The instance has \(2v+5\) jobs:
\begin{itemize}
\item \(v\) \emph{bank jobs of type I}, with period \(1\) and ascent bound \(1\);
\item \(v\) \emph{bank jobs of type II}, with period \(1\) and ascent bound \(2\);
\item one \emph{checker}, with period \(1\) and ascent bound \(G\);
\item two \emph{companions}, with period \(2\) and ascent bound \(2G\);
\item two \emph{fillers}, with period \(2\) and ascent bound \(W\).
\end{itemize}
The intended meaning is that each strip holds one bank job of each type, and that the variable of the strip is true when the type-II job is at the high hour.

\paragraph{Ports.}
The calendar contains two \emph{read phases}. In the read phase of \(F(\mathsf P,\mathsf Q)\), where \((\mathsf P,\mathsf Q)\) is \((\mathsf A,\mathsf B)\) or \((\mathsf B,\mathsf A)\), the current copy \(X\) lives in bank \(\mathsf P\) and the next copy \(Y\) in bank \(\mathsf Q\). The \emph{port} of a literal on \(x_k\) or \(y_k\) is an hour of the strip of variable \(k\) in bank \(\mathsf P\) or \(\mathsf Q\), respectively: the high hour for a positive literal and the low hour for a negative literal. Thus the type-II job occupies the port of a literal exactly when the literal is true. The three literals of a clause are on different variables, so their ports lie in three different strips.

\paragraph{Calendar.}
Let \(O\) denote the set of all \(2v\) ordinary hours. The calendar is a sequence of \emph{blocks} of four days. Within a block we call the days \emph{rows} \(0,1,2,3\). Blocks start on days divisible by \(4\), so rows \(0\) and \(2\) are even days.
\begin{itemize}
\item A \emph{clause block} for a clause with ports \(u_1,u_2,u_3\) offers the hours in \cref{tbl:clause-rows}.
\item A \emph{write phase} for bank \(\mathsf P\) consists of four consecutive \emph{neutral blocks}, that is, \(16\) days \(f=0,\ldots,15\). On day \(f\), every strip of bank \(\mathsf P\) offers \(\{o_\sigma,\,o_\sigma+\omega_f\}\), where
\[
(\omega_0,\ldots,\omega_{15})=(6,1,2,3,4,5,6,8,1,2,3,4,5,6,6,6),
\]
and every other strip offers its ordinary hours. In addition, row \(0\) of each neutral block offers \(p_1-G,p_2-G,p_3-G\), rows \(1\) and \(3\) offer \(p_1,p_2,p_3\), and row \(2\) offers \(p_1+G,p_2+G,p_3+G\). In other words, a neutral block treats the parking hours as its three ports.
\item One calendar period is
\[
\boxed{\text{Read }F(\mathsf A,\mathsf B)}\ \longrightarrow\
\boxed{\text{Write }\mathsf A}\ \longrightarrow\
\boxed{\text{Read }F(\mathsf B,\mathsf A)}\ \longrightarrow\
\boxed{\text{Write }\mathsf B},
\]
where each read phase consists of one clause block for each of \(C_1,\ldots,C_m\). Hence \(M=4m+16+4m+16=4(2m+8)\).
\end{itemize}

\begin{table}[t]
\centering
\small
\begin{tabularx}{\linewidth}{@{}l l X@{}}
\toprule
Row & Available hours & Role \\
\midrule
\(0\) & \(O\cup\{u_1-G,\ u_2-G,\ u_3-G\}\) & Lower guards \\
\(1\) & \(O\cup\{u_1+2,\ u_2+2,\ u_3+2\}\) & Literal offers in the clause's strips \\
\(2\) & \(O\cup\{u_1+G,\ u_2+G,\ u_3+G\}\) & Upper guards \\
\(3\) & \(O\cup\{p_1,\ p_2,\ p_3\}\) & Parking \\
\bottomrule
\end{tabularx}
\caption{The rows of a clause block with ports \(u_1,u_2,u_3\).}\label{tbl:clause-rows}
\end{table}

Every row has \(2v+3=4N+3\) available hours: two per strip and three more. The demand per day is \(2v+1\) jobs of period one plus \(4/2=2\), which is also \(2v+3\). Hence the instance is saturated. All periods divide \(M\). The construction clearly takes polynomial time.

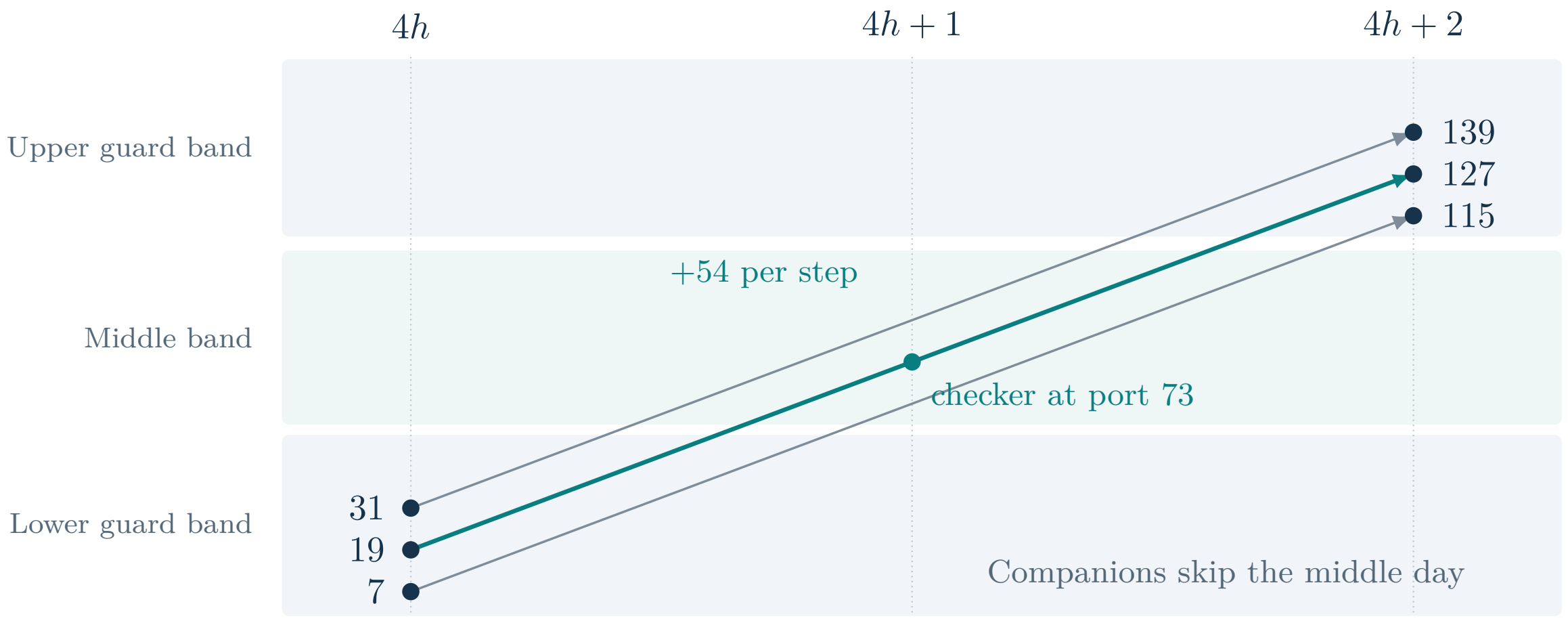
\begin{figure}[t]
\centering
\begin{tikzpicture}[x=1cm,y=.034cm,figuretext]
  \fill[bluefig!7,rounded corners=2pt] (-1.26,0) rectangle (11.25,52);
  \fill[tealfig!7,rounded corners=2pt] (-1.26,55) rectangle (11.25,105);
  \fill[bluefig!7,rounded corners=2pt] (-1.26,109) rectangle (11.25,160);
  \node[anchor=east,font=\scriptsize,text=muted] at (-1.41,26) {Lower guard band};
  \node[anchor=east,font=\scriptsize,text=muted] at (-1.41,80) {Middle band};
  \node[anchor=east,font=\scriptsize,text=muted] at (-1.41,134) {Upper guard band};
  \foreach \x/\t in {0/{4h},4.9/{4h+1},9.8/{4h+2}} {
    \draw[muted!38,densely dotted] (\x,1) -- (\x,161);
    \node[above] at (\x,162) {$\t$};
  }
  \draw[flow,draw=muted!75] (0,7) -- (9.8,115);
  \draw[flow,draw=muted!75] (0,31) -- (9.8,139);
  \draw[flow,draw=tealfig,line width=1.2pt] (0,19) -- (4.9,73) -- (9.8,127);
  \foreach \u in {7,19,31} {
    \node[point] at (0,\u) {};
    \node[anchor=east] at (-.12,\u) {$\u$};
  }
  \foreach \u in {115,127,139} {
    \node[point] at (9.8,\u) {};
    \node[anchor=west] at (9.94,\u) {$\u$};
  }
  \node[point,fill=tealfig] at (4.9,73) {};
  \node[anchor=north west,text=tealfig,font=\footnotesize] at (4.95,71) {checker at port $73$};
  \node[anchor=west,font=\footnotesize,text=muted] at (5.5,12) {Companions skip the middle day};
  \node[anchor=west,font=\footnotesize,text=tealfig] at (2.4,98) {$+54$ per step};
\end{tikzpicture}
\caption{A clause block (\(N=2\), \(v=4\), \(W=50\), \(G=54\)) for the clause \(x_1\vee y_1\vee x_2\), read as \(F(\mathsf A,\mathsf B)\). Its ports are the high hours \(61\), \(85\), and \(73\). The checker uses port \(73\), which requires \(x_2\) to be true, and the bank job at \(73\) then rises to \(75\) in the middle day. The companions rise by \(2G=108\) between consecutive visits. Vertical positions are actual hours; bank jobs and fillers are omitted.}\label{fig:clause}
\end{figure}

\subsection{Completeness}\label{sec:masked-complete}

\begin{lemma}\label{lem:masked-complete}
If \(F\) is satisfiable, the instance with breaks is feasible.
\end{lemma}
\begin{proof}
Take a periodic two-sided satisfying sequence \((X_t)_{t\in\Z}\) of even period; if its period is odd, use two periods. In the \(q\)-th repetition of the calendar, bank \(\mathsf A\) holds \(X_{2q}\) and bank \(\mathsf B\) holds \(X_{2q+1}\) at the start. The write of \(\mathsf A\) stores \(X_{2q+2}\), and the write of \(\mathsf B\) stores \(X_{2q+3}\). Hence the first read phase checks \(F(X_{2q},X_{2q+1})\) and the second checks \(F(X_{2q+1},X_{2q+2})\). We now place the jobs. A strip whose variable has value \(1\) holds its type-II job at the high hour and its type-I job at the low hour; for value \(0\) the two are exchanged.

\emph{Clause block.} Fix a true literal of the clause, with port \(u^*\) in strip \(\sigma^*\), and let \(u',u''\) be the other two ports. The checker uses \(u^*-G\), \(u^*\), \(u^*+G\), \(p_1\) in rows \(0,1,2,3\). The companions are due in rows \(0\) and \(2\); one uses \(u'-G\) and \(u'+G\), and the other \(u''-G\) and \(u''+G\). In row \(1\), the type-II job of strip \(\sigma^*\), which sits at \(u^*\) because the literal is true, moves to \(u^*+2\), and the checker takes \(u^*\). The fillers, due in rows \(1\) and \(3\), take \(u'+2\) and \(u''+2\) in row \(1\) and \(p_2,p_3\) in row \(3\). All other bank jobs stay in place. The ascents are: \(G\) twice for the checker, \(2G\) for the companions, \(2\) for the type-II job, and at most \(W\) for a filler moving inside the middle band. All other moves are descents or zero.

\emph{Neutral block.} The checker uses \(p_1-G,p_1,p_1+G,p_1\); the companions use \(p_2\pm G\) and \(p_3\pm G\); the fillers use \(p_2,p_3\) in rows \(1\) and \(3\). 
On day \(0\) the strip keeps its old value. On days \(1,\ldots,7\) its type-II job takes the upper offered hour. On day \(1\) this is a descent from \(o_\sigma+6\) or an ascent of one from \(o_\sigma\), and on days \(2,\ldots,7\) it is an ascent of at most two. On days \(8,\ldots,15\) the job that represents the new value on top takes the upper hour: on day \(8\) it reaches \(o_\sigma+1\) by an ascent of at most one or by a descent, and it then climbs by one per day. The other job sits at \(o_\sigma\). The strips of the other bank stay in place.

Between blocks, the checker and the companions descend from upper guards to parking hours or lower guards, and the fillers move within the middle band. Every row is exactly filled, every job respects its period, and every ascent bound holds. Hence the schedule is feasible.
\end{proof}

\subsection{Soundness}\label{sec:masked-sound}

Fix a feasible schedule. By \cref{lem:frames} we may assume that it is periodic, and every available hour of every day is occupied.

\begin{lemma}\label{lem:confine}
Every bank job stays inside a single strip \(R_\sigma\) forever, and every strip contains exactly two bank jobs. On every day on which a strip offers exactly two hours, its two bank jobs occupy them.
\end{lemma}
\begin{proof}
Every available hour of every row lies in a region. A bank job has period one and ascent bound at most two, and an upward move from one region to a higher region costs at least four. Hence the region of a bank job never moves up. Since the schedule is periodic, the region never moves down either, so it is constant. The lower guard band is available only in rows \(0\), the upper guard band only in rows \(2\), and the parking set only in rows \(1\) and \(3\). A job of period one needs an hour on every day, so its region is a strip. In row \(0\) of any block, every strip offers exactly two hours. Hence each strip contains at most two bank jobs, and since there are \(2v\) bank jobs and \(v\) strips, each strip contains exactly two. The last claim follows.
\end{proof}

\begin{lemma}\label{lem:phases}
The two companions are due on the even days and the two fillers on the odd days. In rows \(0\) and \(2\) of every block, the checker and the two companions occupy the three guard hours of that row.
\end{lemma}
\begin{proof}
Each day offers \(2v+3\) hours, and \(2v+1\) jobs of period one are due on every day. Since every available hour is occupied, exactly two period-two jobs are due on every day, so two of them use the even phase and two the odd phase. In rows \(0\) and \(2\), every strip offers exactly two hours: in a clause block these are the ordinary hours, and in a write the waveform also offers two hours per strip. By \cref{lem:confine}, the bank jobs fill all strip hours there. So the three remaining hours, which are the guards of the row, hold the checker and the two even period-two jobs. A filler on the even phase would therefore move from a lower guard \(u-G\) in row \(0\) to an upper guard \(u'+G\) in row \(2\), with \(u,u'\) in the middle band. Its ascent would be \(u'-u+2G\ge2G-W>W\), which exceeds its bound. Hence both even jobs are companions.
\end{proof}

\begin{lemma}\label{lem:checker}
Consider a clause block or a neutral block with ports \(u_1,u_2,u_3\). In row \(1\) the checker occupies one of the ports, say \(u_a\), and in rows \(0\) and \(2\) it occupies \(u_a-G\) and \(u_a+G\).
\end{lemma}
\begin{proof}
By \cref{lem:phases}, the checker and the companions occupy the three lower guards in row \(0\) and the three upper guards in row \(2\). Write the checker's hours as \(u_a-G\), \(z\), \(u_b+G\), and let the companions move from \(u_{a'}-G\) to \(u_{b'}+G\) and from \(u_{a''}-G\) to \(u_{b''}+G\), where \(\{a,a',a''\}=\{b,b',b''\}=\{1,2,3\}\). The ascent bound \(G\) of the checker gives \(z\le u_a\) and \(u_b\le z\). The ascent bound \(2G\) of the companions gives \(u_{b'}\le u_{a'}\) and \(u_{b''}\le u_{a''}\). Adding \(u_b\le u_a\), \(u_{b'}\le u_{a'}\), and \(u_{b''}\le u_{a''}\) gives \(u_1+u_2+u_3\le u_1+u_2+u_3\), so all three are equalities. Hence \(u_b=u_a\), so \(z=u_a\).
\end{proof}

\Cref{fig:clause} shows the matching in a concrete clause block.

\begin{lemma}\label{lem:read}
Consider a clause block. (a) The bank job that occupies the checker's port in row \(0\) has type II. (b) In every strip, the bank job at the low hour is the same in rows \(0\), \(2\), and \(3\) of the block and in row \(0\) of the next block.
\end{lemma}
\begin{proof}
(a) Let the checker use port \(u_a\) of strip \(\sigma\) in row \(1\) (\cref{lem:checker}). In row \(1\), strip \(\sigma\) offers \(o_\sigma\), \(o_\sigma+6\), and \(u_a+2\). The checker takes \(u_a\), so the two bank jobs of the strip take the other two hours (\cref{lem:confine}). If \(u_a=o_\sigma\), they take \(o_\sigma+6\) and \(o_\sigma+2\). The job that was at \(o_\sigma\) in row \(0\) cannot reach \(o_\sigma+6\), so it moves to \(o_\sigma+2\); this ascent of two shows that it has type II. If \(u_a=o_\sigma+6\), they take \(o_\sigma\) and \(o_\sigma+8\). The job at \(o_\sigma\) cannot reach \(o_\sigma+8\), so the job at \(o_\sigma+6\) moves to \(o_\sigma+8\) and has type II.

(b) In row \(1\), the job that was at \(o_\sigma\) in row \(0\) is at an hour in \(R_\sigma\) no later than \(o_\sigma+2\), because the only offered hours of \(R_\sigma\) within its reach are \(o_\sigma\) and possibly \(o_\sigma+2\). In row \(2\) the strip offers \(o_\sigma\) and \(o_\sigma+6\), and an hour of at most \(o_\sigma+2\) cannot reach \(o_\sigma+6\). Hence the same job is at \(o_\sigma\) in row \(2\). Rows \(2\), \(3\), and row \(0\) of the next block offer only the ordinary hours in this strip, unless the next block is a write of this strip's bank, whose day \(0\) also offers the ordinary hours. In each case the job at \(o_\sigma\) cannot rise to \(o_\sigma+6\).
\end{proof}

\begin{lemma}\label{lem:write}
(a) During a write of bank \(\mathsf P\), every strip of the other bank keeps the same job at its low hour. (b) Every strip contains exactly one bank job of each type.
\end{lemma}
\begin{proof}
(a) The strips of the other bank offer only their ordinary hours throughout the write, so the job at the low hour never rises to the high hour.

(b) Consider a strip of bank \(\mathsf P\) during the write of \(\mathsf P\). On day \(6\) it offers \(o_\sigma\) and \(o_\sigma+6\), and on day \(7\) it offers \(o_\sigma\) and \(o_\sigma+8\). The job at \(o_\sigma\) on day \(6\) cannot reach \(o_\sigma+8\), so the job at \(o_\sigma+6\) moves to \(o_\sigma+8\), which requires ascent bound two. Hence every strip of \(\mathsf P\) contains a type-II job. Every strip belongs to the active bank of one of the two writes of a calendar period, and its pair of jobs never changes by \cref{lem:confine}. So each of the \(v\) strips contains a type-II job. Since there are exactly \(v\) type-II jobs, each strip contains exactly one, and its other job has type I.
\end{proof}

\begin{lemma}\label{lem:masked-sound}
If the instance with breaks is feasible, then \(F\) is satisfiable.
\end{lemma}
\begin{proof}
By \cref{lem:write}(b), every strip holds one job of each type, so on a day on which a strip offers its ordinary hours, the strip encodes the bit \(1\) if its type-II job is at the high hour and \(0\) otherwise. By \cref{lem:read}(b) and \cref{lem:write}(a), the bits of a bank do not change except during the write of that bank. Number the calendar repetitions by \(q\in\Z\). Let \(X_{2q}\) be the content of bank \(\mathsf A\) and \(X_{2q+1}\) the content of bank \(\mathsf B\) during the first read phase of repetition \(q\). Then bank \(\mathsf A\) holds \(X_{2q+2}\) during the second read phase of repetition \(q\), and bank \(\mathsf B\) still holds \(X_{2q+1}\) there. In every clause block, \cref{lem:read}(a) shows that the job at the checker's port has type II, so the literal of that port is true. Hence the first read phase certifies \(F(X_{2q},X_{2q+1})\) and the second certifies \(F(X_{2q+1},X_{2q+2})\). Thus \((X_t)_{t\in\Z}\) satisfies \(F\).
\end{proof}

\begin{proof}[Proof of \cref{thm:masked}]
The size claims hold by construction, completeness is \cref{lem:masked-complete}, and soundness is \cref{lem:masked-sound}.
\end{proof}

\begin{example}
Consider a relation on two bits whose allowed transitions are \(00\to01\to10\to00\). Start with \((\mathsf A,\mathsf B)=(00,01)\). The first read checks \(00\to01\). The write of \(\mathsf A\) changes it to \(10\) while \(\mathsf B=01\) stays fixed. The second read checks \(01\to10\). The write of \(\mathsf B\) changes it to \(00\), and the next repetition reads \(10\to00\). The same bank jobs store different assignments in different repetitions of the same calendar.
\end{example}

\section{From Daily Scheduling with Breaks to Daily Scheduling}\label{sec:compiler}

\begin{theorem}\label{thm:compiler}
There is a polynomial-time reduction from saturated Daily Scheduling with Breaks to saturated Daily Scheduling. On an instance with width \(w\), calendar period \(M\), and \(n\) jobs, it runs in time polynomial in \(M\), \(w\), and the input length, and it outputs an explicit list of jobs.
\end{theorem}

Fix a saturated instance \(\mathcal M\) with breaks, of width \(w\), calendar period \(M\), rows \(S_0,\ldots,S_{M-1}\), and jobs \((b_i,\delta_i)_{i\in[n]}\) with \(b_i\mid M\). If \(M=1\), we replace the calendar by two copies of its row, so we may assume \(M\ge2\). We construct an instance \(\mathcal U\) without breaks. \Cref{sec:compiler-construction} describes \(\mathcal U\). \Cref{sec:structure,sec:clean,sec:tests,sec:order,sec:timer,sec:pair} establish the structure of every feasible schedule of \(\mathcal U\). \Cref{sec:projection} proves soundness, \cref{sec:compiler-complete} proves completeness, and \cref{sec:compiler-size} bounds the size.

\subsection{The construction}\label{sec:compiler-construction}

\paragraph{Logical slots and labels.}
A \emph{logical slot} is a pair \(j=(t,x)\) with \(0\le t<M\) and \(x\in S_t\); let \(\mathcal S\) be their set. Let
\[
\pv=4M,\qquad E=2M-1 .
\]
Each logical slot \(j=(t,x)\) has two \emph{copies}, with labels \(c_{j,0}=\pv x+2t\) and \(c_{j,1}=\pv x+2t+1\). These labels lie in \(\{\pv,\ldots,\pv w+E\}\), and distinct copies have distinct labels. The \emph{label sequence} is
\[
\lam=(1,2,\ldots,\pv w+E,\ \lambda_0,\lambda_1,\ldots,\lambda_{M-1}),
\]
where \(\lambda_0,\ldots,\lambda_{M-1}\) are \emph{timer labels}. The width of \(\mathcal U\) is \(w'=|\lam|=\pv w+E+M\). The intended schedule puts the \(i\)-th label of \(\lam\) at hour \(i\). Let \(\lam_{\mathrm{app}}\) be the set of copy labels, and \(\lam_{\mathrm{row}}=\lam_{\mathrm{app}}\cup\{\lambda_0,\ldots,\lambda_{M-1}\}\). A copy label \(c\) of a slot in row \(t\) has \emph{row} \(t\), and \(\lambda_t\) has row \(t\).

\paragraph{Groups.}
Each label will carry a few \emph{leaves}, which are reservations of classes of days at its hour. The leaves are partitioned into \emph{groups}:
\begin{itemize}
\item a \emph{layout group} for every two consecutive labels of \(\lam\), with one leaf at each of the two labels;
\item an \emph{equality group} for every copy label \(c\) of row \(t\), with one leaf at \(c\) and one at \(\lambda_t\);
\item the \emph{timer group}, with one leaf at each \(\lambda_t\);
\item the \emph{application group}, with one leaf at each copy label.
\end{itemize}
Thus label \(\ell\) carries one or two layout leaves, and every label of \(\lam_{\mathrm{row}}\) carries additional \emph{row leaves}: an equality leaf and an application leaf at a copy label, and \(2|S_t|\) equality leaves and a timer leaf at \(\lambda_t\). Let \(N_{\mathrm{leaf}}\) be the total number of leaves.

\paragraph{Primes.}
All of the following primes are distinct and larger than \(\max(M,11)\):
\begin{itemize}
\item a \emph{hour prime} \(\pi_\ell\) for every label \(\ell\);
\item an \emph{outer selector} \(s_0\ge3\), a \emph{layout selector} \(s_{\mathrm{lay}}>w'\), and a \emph{row selector} \(s_{\mathrm{row}}>|\lam_{\mathrm{app}}|+2\);
\item two \emph{branch primes} \(h_{\mathrm{lay}}\) and \(h_{\mathrm{row}}\);
\item a \emph{group prime} \(h_g\) for every group \(g\);
\item a \emph{pair prime} \(q_j\) for every logical slot \(j\).
\end{itemize}
Let \(K=2\cdot3\cdot5\cdot7\cdot11=2310\), \(J=Mw^n+2M+10\), and
\[
\Theta=N_{\mathrm{leaf}}\bigl(5(6K+1)+(3M+1)+n(J+2M+3)\bigr).
\]
For every group \(g\), let \(a_g\) be the least integer with \(H_g=h_g^{a_g}>\Theta\). The prime power \(H_g\) is the \emph{group power} of \(g\).

\begin{figure}[t]
\centering
\begin{tikzpicture}[x=1cm,y=1cm, every node/.style={font=\small}]
  \node[box,text width=5.2cm] (rank) at (0,0) {One labelled rank\\ aperture mod \(\pi_\ell\)};
  \node[box,text width=3.6cm,fill=bluefig!7] (lay) at (-3.8,-1.35) {Layout reservation};
  \node[box,text width=5.7cm,fill=tealfig!7] (row) at (2.15,-1.35) {Row reservation and fixed residue \(\tau_\ell\bmod M\)};
  \node[box,text width=3.6cm] (edge) at (-3.8,-2.82) {One leaf per incident\\ adjacency test};
  \node[box,text width=2.45cm] (eq) at (.05,-2.82) {Equality-test\\ leaves};
  \node[box,text width=2.85cm] (app) at (4.25,-2.82) {Application leaf\\ or timer leaf};
  \draw[flow] (rank.south west) -- (lay.north);
  \draw[flow] (rank.south east) -- (row.north);
  \draw[flow] (lay.south) -- (edge.north);
  \draw[flow] (row.south west) -- (eq.north);
  \draw[flow] (row.south east) -- (app.north);
\end{tikzpicture}
\caption{Reservations on a labelled physical rank. A selector gives different time classes to siblings, and a private prime power identifies each leaf group. Equality testing and application are separate descendants of the same row and therefore inherit the same row residue. Layout-only ranks omit the row branch.}\label{fig:reservation-tree}
\end{figure}
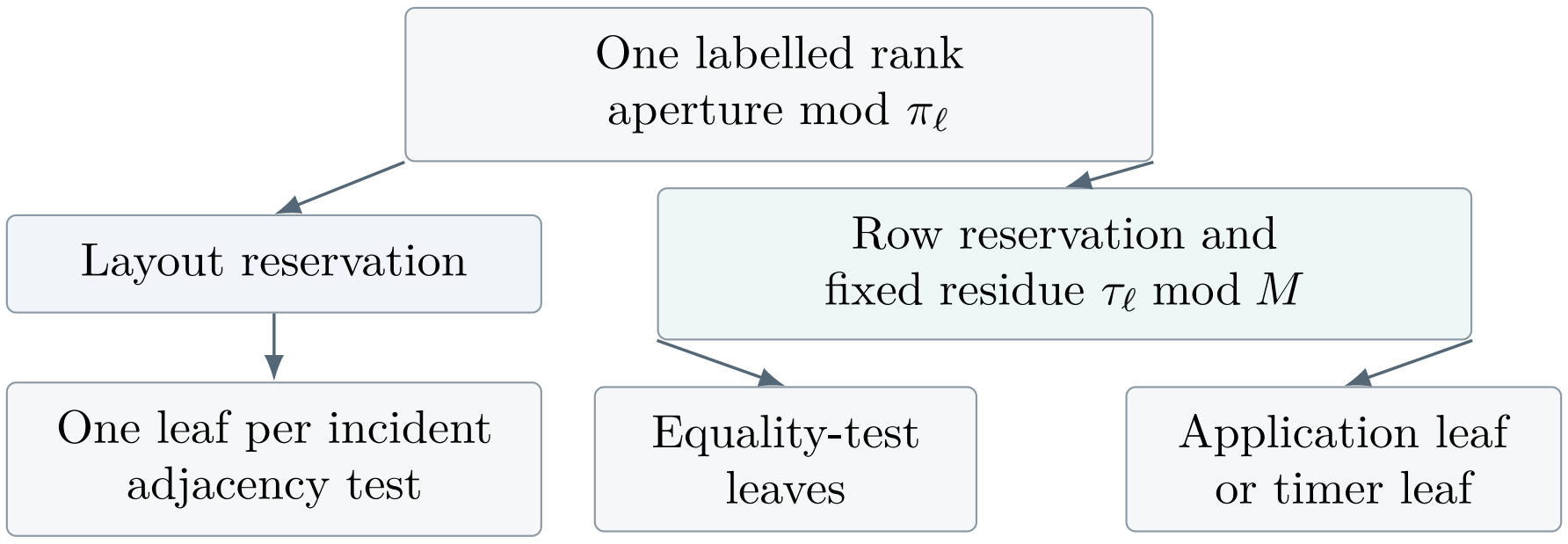

\paragraph{The reservation tree at a label.}
The following rigid jobs carve the hour of label \(\ell\) into its leaves. They are the \emph{structural jobs}. A \emph{split} of a parent modulus \(Q\) with selector \(s\) into children with prime powers \(h_1^{a_1},\ldots,h_k^{a_k}\), where \(k<s\), consists of
\begin{itemize}
\item \(s-k\) \emph{guards} of period \(sQ\), and
\item for each child \(e\in[k]\), a \emph{chain} of \(h_e-1\) jobs of period \(sQh_e^{i}\) for each \(i=1,\ldots,a_e\).
\end{itemize}
The modulus of child \(e\) is \(sQh_e^{a_e}\). At label \(\ell\) the tree is as follows (\cref{fig:reservation-tree}).
\begin{enumerate}[label=(\arabic*)]
\item \emph{Labelling:} \(\pi_\ell-1\) rigid jobs of period \(\pi_\ell\). The root modulus is \(\pi_\ell\).
\item \emph{Outer split} of \(\pi_\ell\) with selector \(s_0\) into a \emph{layout reservation} with prime power \(h_{\mathrm{lay}}\) and, if \(\ell\in\lam_{\mathrm{row}}\), a \emph{row reservation} with prime power \(h_{\mathrm{row}}\).
\item \emph{Row narrowing} (only for \(\ell\in\lam_{\mathrm{row}}\)): \(M-1\) rigid jobs of period \(MQ_\ell\), where \(Q_\ell=s_0\pi_\ell h_{\mathrm{row}}\) is the modulus of the row reservation. The narrowed row reservation has modulus \(MQ_\ell\).
\item \emph{Layout split} of the layout reservation, of modulus \(s_0\pi_\ell h_{\mathrm{lay}}\), with selector \(s_{\mathrm{lay}}\) into the layout leaves of \(\ell\), the leaf of group \(g\) having prime power \(H_g\).
\item \emph{Row split} of the narrowed row reservation with selector \(s_{\mathrm{row}}\) into the row leaves of \(\ell\), the leaf of group \(g\) having prime power \(H_g\).
\end{enumerate}
Hence a leaf of group \(g\) at label \(\ell\) has modulus
\[
P=\pi_\ell s_0h_{\mathrm{lay}}s_{\mathrm{lay}}H_g\ \ \text{(layout leaf)},\qquad
P=\pi_\ell s_0h_{\mathrm{row}}Ms_{\mathrm{row}}H_g\ \ \text{(row leaf)}.
\]
The selector bounds guarantee \(k<s\) at every split.

\paragraph{Contents of the groups.}
The remaining jobs fill the leaves.
\begin{itemize}
\item \emph{Two-hour tests.} Let \(g\) be a layout or equality group with leaves of moduli \(P\) at label \(\ell\) and \(P'\) at label \(\ell'\), and let \(d_g=\lcm(P,P')\). Then \(d_g=P\pi_{\ell'}=P'\pi_\ell\). The group contains \(\pi_{\ell'}-1\) rigid \emph{refiners} of period \(d_g\) for the first leaf and \(\pi_\ell-1\) for the second; five \emph{witnesses} \(\omega_0,\ldots,\omega_4\) of periods \(d_gc_v\), where \((c_0,\ldots,c_4)=(14,33,10,21,55)\); and \(2K-\sum_vK/c_v=4002\) rigid \emph{fillers} of period \(d_gK\). The witnesses have ascent bound \(1\) in a layout group and \(w'-1\) in an equality group.
\item \emph{Timer.} Let \(C_t=\pi_{\lambda_t}s_0h_{\mathrm{row}}s_{\mathrm{row}}H_{\mathrm{tim}}\), so that the timer leaf at \(\lambda_t\) has modulus \(MC_t\), and let \(C=\lcm_tC_t=s_0h_{\mathrm{row}}s_{\mathrm{row}}H_{\mathrm{tim}}\prod_t\pi_{\lambda_t}\). For every \(t\), list the primes \(\pi_{\lambda_{t'}}\) with \(t'\ne t\) as \(c_1,\ldots,c_{M-1}\) and add \(c_h-1\) rigid \emph{timer refiners} of period \(MC_tc_1\cdots c_h\) for each \(h\). Add one \emph{timer job} \(\theta\) of period \(C\) and ascent bound \(1\).
\item \emph{Application.} Let \(B_0=s_0h_{\mathrm{row}}s_{\mathrm{row}}H_{\mathrm{app}}\prod_{c\in\lam_{\mathrm{app}}}\pi_c\). Since \(B_0\) and \(C\) are coprime to \(M\), there is \(m\in[M]\) with \(mB_0\equiv C\pmod M\); it is a unit modulo \(M\). Let \(B=mB_0\) and \(D=MB\). For every logical slot \(j\), with copy labels \(c_0,c_1\), application leaves \(A_{j,0},A_{j,1}\), and their moduli \(P_0,P_1\), the \emph{pair family} \(\mathcal F_j\) consists of:
  \begin{itemize}
  \item two \emph{markers} \(\mu_{j,0},\mu_{j,1}\) of periods \(q_jP_0\) and \(q_jP_1\);
  \item for each side \(s\in\{0,1\}\), writing \(D/P_s=c_1\cdots c_k\) as a product of primes in any order, \((q_j-1)(c_h-1)\) \emph{side refiners} of period \(q_jP_sc_1\cdots c_h\) for each \(h\in[k]\);
  \item \(q_j-2\) \emph{pair refiners} of period \(q_jD\).
  \end{itemize}
  Note that \(D/P_s=m\prod_{c\in\lam_{\mathrm{app}}\setminus\{c_s\}}\pi_c\).
\item \emph{Computation jobs.} For every job \(i\) of \(\mathcal M\), a job of period \(Bb_i\) and ascent bound \(\pv\delta_i+E\).
\end{itemize}
All jobs not declared moving above are rigid. The moving jobs are the witnesses, the timer job, and the computation jobs. \Cref{tbl:families} in \cref{app:notation} lists all families.

\begin{lemma}\label{lem:saturation}
The instance \(\mathcal U\) is saturated.
\end{lemma}
\begin{proof}
We show that the rates \(1/p\) of the jobs assigned to each label sum to one. The labelling jobs have rate \((\pi_\ell-1)/\pi_\ell\) and leave \(1/\pi_\ell\). In a split of a parent of rate \(1/Q\), the guards have rate \((s-k)/(sQ)\), the chain of child \(e\) has rate \((h_e-1)\sum_{i=1}^{a_e}1/(sQh_e^i)=(1-h_e^{-a_e})/(sQ)\), and the child itself has rate \(1/(sQh_e^{a_e})\). These add up to \(1/Q\). The row narrowing has rate \((M-1)/(MQ_\ell)\) and leaves \(1/(MQ_\ell)\). It remains to check that the contents of each group have total rate equal to the sum \(\sum1/P\) over its leaves.

For a two-hour test, the refiners have rate \((\pi_{\ell'}-1)/d_g+(\pi_\ell-1)/d_g=1/P+1/P'-2/d_g\). The witnesses and fillers have rate \(\sum_v1/(d_gc_v)+(2K-\sum_vK/c_v)/(d_gK)=2/d_g\). For the timer, the refiners of the leaf at \(\lambda_t\) have rate \(1/(MC_t)-1/(MC)\), and the timer job has rate \(1/C=M/(MC)\). For a pair family, the markers have rate \((1/P_0+1/P_1)/q_j\), the side refiners \((1-1/q_j)(1/P_s-1/D)\) on side \(s\), and the pair refiners \((q_j-2)/(q_jD)\). Their sum is \(1/P_0+1/P_1-1/D\). Finally, the computation jobs have rate \(\sum_i1/(Bb_i)=\frac1B\cdot\frac1M\sum_t|S_t|=|\mathcal S|/D\), by the saturation of \(\mathcal M\). This supplies the missing \(1/D\) for each of the \(|\mathcal S|\) pair families.
\end{proof}

From now on we fix a feasible schedule of \(\mathcal U\). By \cref{lem:frames} we may assume that it is periodic and that every slot is occupied. By \cref{lem:rigid}, every rigid job has a footprint.

\subsection{The structure forced on rigid jobs}\label{sec:structure}

\begin{lemma}\label{lem:labels}
There is a bijection \(\phi\) from the labels to the hours such that all labelling jobs of \(\ell\) lie at hour \(\phi(\ell)\). They cover all classes modulo \(\pi_\ell\) at that hour except one, \(\ap{\zeta_\ell}{\pi_\ell}\). Every other rigid job at hour \(\phi(\ell)\) has a period divisible by \(\pi_\ell\) and days in \(\ap{\zeta_\ell}{\pi_\ell}\). In particular, a rigid job whose period is divisible by exactly one hour prime \(\pi_\ell\) lies at hour \(\phi(\ell)\).
\end{lemma}
\begin{proof}
Two rigid jobs at one hour whose periods are coprime collide by \cref{lem:rigid}. Hence labelling jobs of different labels never share an hour, and each hour carries labelling jobs of at most one label. There are \(w'\) labels and \(w'\) hours, and every label has at least one labelling job, so each label occupies exactly one hour. The \(\pi_\ell-1\) labelling jobs of \(\ell\) have the same period and are pairwise disjoint, so they occupy \(\pi_\ell-1\) distinct classes modulo \(\pi_\ell\). A rigid job at hour \(\phi(\ell)\) whose period is not divisible by \(\pi_\ell\) meets every class modulo \(\pi_\ell\) and collides. A rigid job whose period is divisible by \(\pi_\ell\) avoids the labelling jobs only if its days lie in the free class.
\end{proof}

The next claim describes one split. Fix an hour \(\phi(\ell)\), a class \(X=\ap xQ\) of days, and a split of modulus \(Q\) with selector \(s\) and children \(e\in[k]\). Assume that the footprints of the guards and chain jobs of this split lie in \(\{\phi(\ell)\}\times X\).

\begin{claim}\label{clm:split}
The guards occupy \(s-k\) distinct classes modulo \(sQ\) inside \(X\). The chain of each child \(e\) lies inside one further class \(\ap{\varsigma_e}{sQ}\), and distinct children use distinct classes. Inside \(\ap{\varsigma_e}{sQ}\), the chain covers every day except a single class \(Y_e\) modulo \(sQh_e^{a_e}\). Consequently, the days of \(X\) not covered by the guards and chains of this split are exactly \(\bigsqcup_eY_e\).
\end{claim}
\begin{proof}
Guards have the same period \(sQ\) and are pairwise disjoint, so they use distinct classes modulo \(sQ\). Each chain job lies in one class modulo \(sQ\), because \(sQ\) divides its period. A chain job of child \(e\) and a guard, or chain jobs of two different children, have periods whose greatest common divisor is \(sQ\), because the child primes are distinct. By \cref{fact:crt}, they collide if they lie in the same class modulo \(sQ\). So the guards and the \(k\) chains use pairwise disjoint sets of classes modulo \(sQ\). There are \(s\) classes, the guards use \(s-k\), and each chain is nonempty, so each chain uses exactly one class \(\ap{\varsigma_e}{sQ}\).

Fix a child \(e\) and write \(h=h_e\), \(a=a_e\). The \(h-1\) chain jobs of period \(sQh\) are disjoint and lie in \(\ap{\varsigma_e}{sQ}\), so they cover \(h-1\) of its \(h\) subclasses modulo \(sQh\), leaving one, \(Y^{(1)}\). Suppose that the jobs of periods \(sQh^{i'}\) for \(i'<i\) cover everything in \(\ap{\varsigma_e}{sQ}\) except one class \(Y^{(i-1)}\) modulo \(sQh^{i-1}\). A job of period \(sQh^i\) lies in one class modulo \(sQh^{i-1}\), and if that class is covered by a job of period \(sQh^{i'}\) with \(i'<i\), the two collide by \cref{fact:crt}. So all jobs of period \(sQh^i\) lie in \(Y^{(i-1)}\), and they cover \(h-1\) of its \(h\) subclasses modulo \(sQh^i\), leaving one, \(Y^{(i)}\). Take \(Y_e=Y^{(a)}\).
\end{proof}

\begin{claim}\label{clm:descent}
In the setting of \cref{clm:split}, let \(J\) be a rigid job, not a guard or chain job of this split, with a slot in \(\{\phi(\ell)\}\times X\) and with \(Q\) dividing its period \(p\). Then all days of \(J\) lie in a single \(Y_e\), and \(sh_e^{a_e}\) divides \(p\).
\end{claim}
\begin{proof}
Since \(Q\mid p\), the days of \(J\) form a class \(Z\) modulo \(p\) contained in \(X\). They avoid the days covered by the split. If \(s\nmid p\), then \(Z\) meets all \(s\) classes modulo \(sQ\) inside \(X\), including a guard class, because \(s-k\ge1\). So \(s\mid p\), and \(Z\) lies in one class modulo \(sQ\), which is not a guard class; let it be \(\ap{\varsigma_e}{sQ}\). Let \(h^c\) be the largest power of \(h=h_e\) dividing \(p\). For \(i\le a_e\), if \(Z\) lies inside \(Y^{(i-1)}\) (where \(Y^{(0)}=\ap{\varsigma_e}{sQ}\)) and \(c<i\), then \(Z\) meets all \(h\) subclasses of \(Y^{(i-1)}\) modulo \(sQh^i\), and \(h-1\) of them are covered by chain jobs, a contradiction. Hence \(c\ge i\), and \(Z\) lies in a single subclass modulo \(sQh^i\), which must be the free one, \(Y^{(i)}\). By induction, \(c\ge a_e\) and \(Z\subseteq Y_e\).
\end{proof}

\begin{claim}\label{clm:narrow}
Let \(X\) be the days of the row reservation at a label \(\ell\in\lam_{\mathrm{row}}\), a class modulo \(Q_\ell\), and assume that the \(M-1\) narrowing jobs of \(\ell\) lie in \(\{\phi(\ell)\}\times X\). They cover all classes modulo \(MQ_\ell\) inside \(X\) except one. Its residue modulo \(M\) is denoted \(\tau_\ell\). Every other rigid job with a slot in \(\{\phi(\ell)\}\times X\) whose period \(p\) is divisible by \(Q_\ell\) satisfies \(M\mid p\) and has all its days in the free class.
\end{claim}
\begin{proof}
The narrowing jobs have the same period and are disjoint, so they cover \(M-1\) of the \(M\) classes modulo \(MQ_\ell\) inside \(X\). Let \(J\) be another such rigid job and \(Z\) its class of days. Since \(\gcd(Q_\ell,M)=1\), the class \(Z\) meets exactly \(M/\gcd(p/Q_\ell,M)\) classes modulo \(MQ_\ell\) inside \(X\), by the remark after \cref{fact:crt}. At most one of them is free, so \(M\mid p\) and \(Z\) lies in the free class.
\end{proof}

We now apply the claims from the root of each tree down to its leaves. A rigid job is \emph{non-structural} if it is not a labelling job, a guard, a chain job, or a narrowing job.

\begin{lemma}\label{lem:structure}
At every label \(\ell\), the structural jobs of \(\ell\) lie at hour \(\phi(\ell)\) and are arranged as in \cref{clm:split,clm:narrow} at every node of the tree. Every leaf of \(\ell\) has an \emph{aperture}: a single class \(A\) of days modulo the leaf modulus \(P\), at hour \(\phi(\ell)\). The slots of hour \(\phi(\ell)\) that lie outside the footprints of structural jobs are exactly the disjoint union of the apertures of the leaves of \(\ell\).
\end{lemma}
\begin{proof}
Every structural job of \(\ell\) has exactly one hour prime in its period, namely \(\pi_\ell\), so it lies at hour \(\phi(\ell)\) with days in \(\ap{\zeta_\ell}{\pi_\ell}\) (\cref{lem:labels}). Apply \cref{clm:split} to the outer split with \(X=\ap{\zeta_\ell}{\pi_\ell}\). The narrowing jobs and the jobs of the inner splits have periods divisible by \(s_0\pi_\ell\) and by exactly one branch prime. By \cref{clm:descent}, the narrowing jobs and the jobs of the row split lie in the row reservation, and the jobs of the layout split lie in the layout reservation. Apply \cref{clm:narrow} to the row reservation. The jobs of the row split have periods divisible by \(MQ_\ell\), so by \cref{clm:narrow} they lie in the narrowed row reservation. Apply \cref{clm:split} to the two inner splits. Each step accounts for every day of its parent, which gives the final claim.
\end{proof}

\begin{lemma}\label{lem:scope}
Every non-structural rigid job \(J\) lies inside the aperture of a single leaf, and the modulus \(P\) of that leaf divides the period of \(J\). In particular, \(J\) lies in a leaf of group \(g\) only if the group prime \(h_g\) divides its period.
\end{lemma}
\begin{proof}
By \cref{lem:labels}, \(J\) lies at some hour \(\phi(\ell)\) with \(\pi_\ell\) dividing its period and its days in \(\ap{\zeta_\ell}{\pi_\ell}\). Apply \cref{clm:descent} to the outer split. If \(J\) lands in the row reservation, apply \cref{clm:narrow} and then \cref{clm:descent} to the row split; otherwise apply \cref{clm:descent} to the layout split. Each application shows that the factor of the next level divides the period of \(J\) and that all days of \(J\) lie in the next node.
\end{proof}

Every family of the construction has periods divisible by the group prime of its own group and by no other group prime. By \cref{lem:scope}, the rigid jobs of each group lie in the leaves of that group. We also record where the markers lie.

\begin{lemma}\label{lem:markers}
For every logical slot \(j\) and side \(s\), the marker \(\mu_{j,s}\) lies in the aperture of the application leaf \(A_{j,s}\). No rigid job of a pair family \(\mathcal F_{j'}\) with \(j'\ne j\) has a slot in \(A_{j,0}\) or \(A_{j,1}\).
\end{lemma}
\begin{proof}
The period \(q_jP_s\) of \(\mu_{j,s}\) has exactly one hour prime, \(\pi_{c_s}\), so \(\mu_{j,s}\) lies at hour \(\phi(c_s)\). By \cref{lem:scope}, it lies in a leaf of \(c_s\) whose modulus divides \(q_jP_s\). The modulus of every other leaf of \(c_s\) contains a layout or equality group prime, which does not divide \(q_jP_s\). So the leaf is \(A_{j,s}\).

Now let \(J\in\mathcal F_{j'}\) with \(j'\ne j\) have a slot in \(A_{j,s}\). By \cref{lem:scope}, its period \(p\) is divisible by \(P_s\), and all its days lie in \(A_{j,s}\). The days of \(J\) form a class modulo \(p\), and those of \(\mu_{j,s}\) a class modulo \(q_jP_s\); both lie in the class \(A_{j,s}\) modulo \(P_s\). Since \(q_j\nmid p\), we have \(\gcd(p,q_jP_s)=P_s\), so the two classes intersect by \cref{fact:crt}, and \(J\) collides with \(\mu_{j,s}\).
\end{proof}

\subsection{Clean windows}\label{sec:clean}

Moving jobs are not confined by their periods. We only need them to stay in their own group for a long but finite interval. A leaf is \emph{in scope} for a set of moving jobs if it belongs to a prescribed set \(\Gamma\) of leaves.

\begin{lemma}\label{lem:clean}
Let \(\mathcal J\) be a set of moving jobs whose periods divide \(X\), and let \(\Gamma\) be a set of leaves such that \(\gcd(X,h_g)=1\) for the group \(g\) of every leaf not in \(\Gamma\). Let \(I\) be a finite interval of days, and let \(N_{\mathrm{due}}\) be the number of days of \(I\) on which a job of \(\mathcal J\) is due. If \(H_g>N_{\mathrm{leaf}}N_{\mathrm{due}}\) for every leaf not in \(\Gamma\), then there are infinitely many integers \(v\), positive and negative, such that in the translated interval \(I+vX\) every job of \(\mathcal J\) occupies, on every day it is due, a slot of an aperture of a leaf in \(\Gamma\). We call such a translate \emph{clean} for \(\mathcal J\).
\end{lemma}
\begin{proof}
Since every period of \(\mathcal J\) divides \(X\), the days of \(I+vX\) on which a job of \(\mathcal J\) is due are the translates \(o+vX\) of the corresponding days \(o\) of \(I\). A moving job occupies a slot outside every rigid footprint, so by \cref{lem:structure} the slot lies in the aperture of some leaf. The aperture of a leaf \(\lambda\notin\Gamma\) of group \(g\) is a class modulo its modulus \(P\), and \(H_g\mid P\), so its days lie in a single class modulo \(H_g\). Since \(\gcd(X,H_g)=1\), the condition \(o+vX\in\) (that class) holds for \(v\) in exactly one class modulo \(H_g\). Excluding these classes for all due days \(o\) and all leaves \(\lambda\notin\Gamma\) removes a set of integers \(v\) of density at most \(N_{\mathrm{leaf}}N_{\mathrm{due}}\max_g1/H_g<1\). The remaining set has positive density in both directions, and for every \(v\) in it, no leaf outside \(\Gamma\) has a slot on any due day of \(I+vX\).
\end{proof}

We use three kinds of clean windows. The exponents \(a_g\) were chosen so that \(H_g>\Theta\) covers all of them.
\begin{itemize}
\item For the witnesses of a two-hour test \(g\): \(X=d_gK\), \(I\) of length \(6d_gK\), \(\Gamma\) the two leaves of \(g\). The witnesses have periods at least \(d_g\), so \(N_{\mathrm{due}}\le5(6K+1)\).
\item For the timer job: \(X=MC\), \(I\) of length \(3MC\), \(\Gamma\) the timer leaves, and \(N_{\mathrm{due}}\le3M+1\).
\item For the computation jobs: \(X=MB\), \(I\) of length \(B(J+2M+2)\), \(\Gamma\) the application leaves, and \(N_{\mathrm{due}}\le n(J+2M+3)\).
\end{itemize}
In each case the prime factors of \(X\) are hour primes, selectors, branch primes, primes at most \(M\), primes at most \(11\), and the group prime of the in-scope group. They do not include the group prime of any other group, so the coprimality condition holds.

\subsection{Two-hour tests}\label{sec:tests}

\begin{figure}[t]
\centering
\begin{minipage}[c]{.37\linewidth}
\centering
\begin{tikzpicture}[every node/.style={font=\small,text=ink}]
  \foreach \v/\angle in {0/90,1/18,2/-54,3/-126,4/162} {
    \node[circle,draw=ink!70,fill=white,minimum size=6.8mm] (v\v) at (\angle:1.55cm) {\(\v\)};
  }
  \draw[tealfig,line width=1.1pt] (v0) -- (v1) -- (v2) -- (v3) -- (v4) -- (v0);
  \draw[muted,dashed] (v0) -- (v2) -- (v4) -- (v1) -- (v3) -- (v0);
\end{tikzpicture}
\end{minipage}\hfill
\begin{minipage}[c]{.59\linewidth}
\small\centering
\begin{tabular}{@{}clccc@{}}
\toprule
Job \(v\) & Nonedge primes & \(c_v\) & \(\beta_v\) \\
\midrule
0 & \(2,7\) & 14 & 0 \\
1 & \(3,11\) & 33 & 0 \\
2 & \(2,5\) & 10 & 5 \\
3 & \(3,7\) & 21 & 1 \\
4 & \(5,11\) & 55 & 1 \\
\bottomrule
\end{tabular}
\end{minipage}
\caption{The five-cycle test. Solid edges join jobs with coprime suffix periods, which must meet when they choose the same phase modulo \(d\). Dashed pairs share one prime. The example uses different suffix phases modulo that prime for the two endpoints of each dashed pair.}\label{fig:test}
\end{figure}

\begin{lemma}\label{lem:tworank}
Let \(g\) be a layout or equality group with leaves at labels \(\ell\ne\ell'\), apertures \(A,A'\), and \(d=d_g\).
\begin{enumerate}[label=(\alph*)]
\item There is a residue \(u\) modulo \(d\) such that \(A\) and \(A'\) both contain slots on every day of \(\ap ud\). In particular, the two apertures have compatible residues modulo every common factor of their moduli.
\item If the witnesses of \(g\) have ascent bound \(1\), then \(|\phi(\ell)-\phi(\ell')|=1\).
\item Conversely, consider any placement of the rigid jobs outside \(g\) in which \(A\) and \(A'\) are classes whose days share a class modulo \(d\), and, in the case of ascent bound \(1\), in which \(|\phi(\ell)-\phi(\ell')|=1\). Then the jobs of \(g\) can be placed so that they occupy exactly the slots of \(A\cup A'\), with a schedule of period \(dK\).
\end{enumerate}
\end{lemma}
\begin{proof}
(a) By \cref{lem:scope}, the rigid jobs with slots in \(A\cup A'\) are the refiners and fillers of \(g\). The days of \(A\) form a class modulo \(P\mid d\), so \(A\) has \(d/P=\pi_{\ell'}\) slots in every \(d\) consecutive days; similarly \(A'\) has \(\pi_\ell\). The refiners have period \(d\), and there are \(\pi_\ell+\pi_{\ell'}-2\) of them. Let \(V_0\) be the set of slots of \(A\cup A'\) not covered by refiners. It is invariant under translation by \(d\) days and contains exactly two slots in every \(d\) consecutive days. Let their days have residues \(u_1,u_2\) modulo~\(d\).

Take a clean translate \(I'\) of length \(6dK\) for the witnesses (\cref{lem:clean}). Inside \(I'\), every witness occupies a slot of \(V_0\) on every day it is due. Witness \(\omega_v\) is due only on days of a fixed class \(\alpha_v\) modulo \(d\), because \(d\) divides its period, so \(\alpha_v\in\{u_1,u_2\}\). Suppose that \(u_1\ne u_2\). Then every day contains at most one slot of \(V_0\). For adjacent vertices \(v,v'\) of the five-cycle, \(\gcd(c_v,c_{v'})=1\) (\cref{fig:test}), so \(\gcd(dc_v,dc_{v'})=d\). If \(\alpha_v=\alpha_{v'}\), the days on which both are due form a class modulo \(dc_vc_{v'}\), by \cref{fact:crt}, and \(dc_vc_{v'}\) divides \(dK\). So both are due on some day of \(I'\), which has only one slot of \(V_0\). Hence adjacent witnesses have different residues, and \(v\mapsto\alpha_v\) is a proper two-colouring of the five-cycle, which is impossible. Therefore \(u_1=u_2=u\). The two slots of \(V_0\) on a day of \(\ap ud\) are at different hours, so one lies in \(A\) and one in \(A'\).

(b) Suppose that \(|\phi(\ell)-\phi(\ell')|\ge2\) and the ascent bound is \(1\). By (a), inside \(I'\) every witness occupies one of the hours \(\phi(\ell)\) and \(\phi(\ell')\) whenever it is due, and it can never move from the lower of the two to the higher. So each witness changes hour at most once in \(I'\). Split \(I'\) into six consecutive intervals of length \(dK\), and mark, for each witness that changes hour, the interval containing the later visit of its change. Some interval is unmarked, and in it every witness keeps one hour. All witnesses have the same residue \(u\) modulo \(d\), so adjacent witnesses are both due on some day of every interval of length \(dK\). On that day they need different hours. This gives a proper two-colouring of the five-cycle by hour, which is impossible.

(c) Let \(\ap ud\) be the common class. Place the refiners of each leaf on the \(d/P-1\) classes modulo \(d\) of its aperture other than \(\ap ud\). Give witness \(\omega_v\) the phase \(u+d\beta_v\) modulo \(dc_v\), with \((\beta_0,\ldots,\beta_4)=(0,0,5,1,1)\) as in \cref{fig:test}. The two endpoints of every non-edge of the five-cycle share exactly one prime \(p\) among their \(c_v\), and their values \(\beta_v\) differ modulo \(p\), so they are never due together. Hence the witnesses due on one day form a clique of the five-cycle and number at most two. On each day, place the due witnesses at the two hours by a fixed rule that depends only on the day modulo \(dK\). Every move is between \(\phi(\ell)\) and \(\phi(\ell')\), which is allowed by ascent bound \(w'-1\), and by ascent bound \(1\) when the hours are adjacent. In every \(dK\) consecutive days there are \(2K\) slots of \(V_0\), of which the witnesses use \(\sum_vK/c_v=618\). Place one filler, of period \(dK\), on each of the remaining \(4002\) slots.
\end{proof}

\subsection{Order of the labels and row residues}\label{sec:order}

\begin{lemma}\label{lem:layout}
Either \(\phi(\lam_i)=i\) for every \(i\), or \(\phi(\lam_i)=w'+1-i\) for every \(i\), where \(\lam_i\) is the \(i\)-th label of \(\lam\).
\end{lemma}
\begin{proof}
By \cref{lem:tworank}(b) applied to the layout groups, \(|\phi(\lam_{i+1})-\phi(\lam_i)|=1\) for every \(i<w'\). A sequence of \(w'\) distinct elements of \([w']\) whose consecutive elements differ by one cannot change direction, since it would then repeat a value. So it is \(1,2,\ldots,w'\) or \(w',w'-1,\ldots,1\).
\end{proof}

We write \(\varepsilon=1\) in the first case and \(\varepsilon=-1\) in the second. Recall the row residues \(\tau_\ell\) from \cref{clm:narrow}.

\begin{lemma}\label{lem:rows}
For every copy label \(c\) of row \(t\), we have \(\tau_c=\tau_{\lambda_t}\). The days of every row leaf at a label \(\ell\) lie in \(\ap{\tau_\ell}M\).
\end{lemma}
\begin{proof}
The second claim holds because every row leaf lies inside the narrowed row reservation. The equality group of \(c\) has leaves at \(c\) and \(\lambda_t\), and \(M\) divides \(d_g\). By \cref{lem:tworank}(a), the two leaves contain a common class modulo \(d_g\), hence a common class modulo \(M\). So \(\tau_c=\tau_{\lambda_t}\).
\end{proof}

\subsection{The timer orders the rows}\label{sec:timer}

\begin{lemma}\label{lem:timer}
There is a residue \(\tau_0\) such that \(\tau_{\lambda_t}\equiv\tau_0+\varepsilon Ct\pmod M\) for every \(t\).
\end{lemma}
\begin{proof}
Take a clean translate \(I'\) of length \(3MC\) for the timer job (\cref{lem:clean}). Inside it, \(\theta\) is due on \(M+1\) consecutive days \(s_0,s_0+C,\ldots,s_0+MC\), and on each it occupies a slot of a timer leaf. Since \(\gcd(C,M)=1\), the days \(s_0+kC\), \(k=0,\ldots,M-1\), have all \(M\) residues modulo \(M\). The timer leaf at \(\lambda_t\) has days only in \(\ap{\tau_{\lambda_t}}M\) by \cref{lem:rows}. So every residue modulo \(M\) equals \(\tau_{\lambda_t}\) for some \(t\), and the map \(t\mapsto\tau_{\lambda_t}\) is a bijection onto \(\Z/M\Z\). Consequently, on day \(s_0+kC\) the timer job is at the hour \(y_k=\phi(\lambda_{t_k})\), where \(t_k\) is the unique index with \(\tau_{\lambda_{t_k}}\equiv s_0+kC\pmod M\), and \(y_M=y_0\).

By \cref{lem:layout}, the timer hours form an interval \(\{b,\ldots,b+M-1\}\), and \(y_0,\ldots,y_{M-1}\) is a permutation of it. Every step satisfies \(y_{k+1}-y_k\le1\). Let \(k_0\) be the index with \(y_{k_0}=b\), and read the indices modulo \(M\). Then \(y_{k_0+1}\le b+1\) and \(y_{k_0+1}\ne b\), because \(M\ge2\) and the values are distinct; so \(y_{k_0+1}=b+1\). Inductively, \(y_{k_0+i}=b+i\) for \(0\le i<M\). Thus the residue at hour \(b+i\) is \(\rho+iC\) with \(\rho=s_0+k_0C\). If \(\varepsilon=1\), label \(\lambda_t\) is at hour \(b+t\) and \(\tau_{\lambda_t}=\rho+Ct\). If \(\varepsilon=-1\), label \(\lambda_t\) is at hour \(b+M-1-t\), and \(\tau_{\lambda_t}=\rho+(M-1)C-Ct\). In both cases the claim holds with \(\tau_0=\tau_{\lambda_0}\).
\end{proof}

\subsection{One vacancy per logical slot}\label{sec:pair}

The leaves of the application group may have different moduli. A naive refinement to a common modulus does not preserve the number of vacancies at each hour, because refiners may migrate between leaves.

\begin{example}\label{ex:migrate}
Consider two apertures \(\ap03\) and \(\ap05\) at two hours, refined to modulus \(15\). The natural refinement adds four jobs of period \(15\) for the first hour and two for the second. The six jobs can occupy the days \(3,6,9\) at the first hour and \(0,5,10\) at the second, modulo \(15\). Then the first hour has two vacancies per period and the second has none.
\end{example}

The pair families avoid this. For a logical slot \(j\), let \(V_j(s)\) be the number of slots of \(A_{j,0}\cup A_{j,1}\) on day \(s\) that lie outside every rigid footprint.

\begin{lemma}\label{lem:pair}
For every logical slot \(j=(t,x)\) there is a residue \(u_j\) modulo \(D\) with \(u_j\equiv\tau_{\lambda_t}\pmod M\) such that \(V_j(s)=1\) if \(s\equiv u_j\pmod D\) and \(V_j(s)=0\) otherwise.
\end{lemma}
\begin{proof}
By \cref{lem:scope,lem:markers}, the rigid jobs with slots in \(A_{j,0}\cup A_{j,1}\) are exactly the members of \(\mathcal F_j\). Indeed, only pair families have the application group prime, families \(\mathcal F_{j'}\) with \(j'\ne j\) are excluded by \cref{lem:markers}, and every member of \(\mathcal F_j\) lies in some application leaf, hence in \(A_{j,0}\) or \(A_{j,1}\) by the same lemma applied to the other slots. Let \(R(s)\) be the number of the two apertures with a slot on day \(s\), and let \(F(s)\) be the number of members of \(\mathcal F_j\) due on day \(s\). Each due member occupies one slot of \(A_{j,0}\cup A_{j,1}\) on its day, so \(V_j(s)=R(s)-F(s)\). Since \(P_0,P_1\mid D\), the function \(R\) depends only on \(s\bmod D\).

We show that \(F\) also depends only on \(s\bmod D\). Let \(L'\) be the least common multiple of the periods of all jobs outside \(\mathcal F_j\). The prime \(q_j\) does not divide \(L'\). The narrowing jobs have periods divisible by \(M\), and the computation jobs have periods divisible by \(B\), so \(D=MB\) divides \(L'\). By \cref{lem:frames}(c), exactly \(w'\) jobs are due on every day. The number of jobs outside \(\mathcal F_j\) due on day \(s\) depends only on \(s\bmod L'\), and hence so does \(F(s)\). On the other hand, every member of \(\mathcal F_j\) has a period dividing \(q_jD\), so \(F(s)\) depends only on \(s\bmod q_jD\). Let \(s\equiv s'\pmod D\). By \cref{fact:crt}, there is \(s''\) with \(s''\equiv s\pmod{L'}\) and \(s''\equiv s'\pmod{q_j}\). Then \(F(s'')=F(s)\). Moreover \(s''\equiv s\equiv s'\pmod D\), so \(s''\equiv s'\pmod{q_jD}\) and \(F(s'')=F(s')\). Hence \(F(s)=F(s')\).

So \(V_j\) is a nonnegative integer function of \(s\bmod D\). Its average is \(1/P_0+1/P_1\) minus the rate of \(\mathcal F_j\), which is \(1/D\) by the computation in the proof of \cref{lem:saturation}. Hence \(V_j\) equals one on exactly one residue \(u_j\) modulo \(D\) and zero elsewhere. Since \(V_j(u_j)=1\), the day \(u_j\) belongs to \(A_{j,0}\) or \(A_{j,1}\), and by \cref{lem:rows} both lie in \(\ap{\tau_{\lambda_t}}M\).
\end{proof}

\begin{figure}[t]
\centering
\begin{tikzpicture}[x=1cm,y=1cm, every node/.style={font=\small,text=ink}]
  \node[anchor=east] at (-.75,1.45) {Private coordinate $t\bmod 5$};
  \node[anchor=east] at (-.75,.7) {Physical copy 0};
  \node[anchor=east] at (-.75,0) {Physical copy 1};
  \foreach \z in {0,...,4} {\node at (\z*1.40,1.45) {$\z$};}
  \foreach \z/\y/\shade/\txt in {0/.7/bluefig!18/marker,1/.7/ochre!19/open,2/.7/tealfig!15/refiner,3/.7/tealfig!15/refiner,4/.7/tealfig!15/refiner,0/0/ochre!19/open,1/0/bluefig!18/marker,2/0/ochre!19/open,3/0/ochre!19/open,4/0/ochre!19/open}{
    \node[draw=ink!40,fill=\shade,minimum width=1.26cm,minimum height=.55cm,rounded corners=1.5pt,font=\footnotesize] at (\z*1.40,\y) {\txt};
  }
  \node[anchor=east,font=\footnotesize\bfseries] at (-.75,-.69) {Vacancies across the pair};
  \foreach \z in {0,...,4} {\node[font=\footnotesize\bfseries,text=ochre] at (\z*1.40,-.69) {1};}
\end{tikzpicture}
\caption{The paired invariant in a small completeness placement. The raw apertures are \(0\bmod 2\) and \(0\bmod 3\), with \(D=6\) and \(q=5\). The markers have periods \(10\) and \(15\). Eight and four period-\(30\) refiners fill the unmarked slices on the other coarse residues. Three more period-\(30\) refiners resolve the double openings shown at coarse residue \(0\bmod 6\). The available physical copy can change with the private coordinate.}\label{fig:paired-example}
\end{figure}

\begin{lemma}\label{lem:pair-complete}
Suppose that \(A_{j,0}\) and \(A_{j,1}\) are classes whose days both contain the class \(\ap{u}D\). Then the members of \(\mathcal F_j\) can be placed in \(A_{j,0}\cup A_{j,1}\) so that, on every day of \(\ap uD\), exactly one slot of \(A_{j,0}\cup A_{j,1}\) remains free, and no slot remains free on other days.
\end{lemma}
\begin{proof}
Write \(q=q_j\). Give the two markers different residues modulo \(q\), so that each covers one of the \(q\) classes modulo \(qP_s\) of its aperture. On side \(s\), the \(q-1\) uncovered classes modulo \(qP_s\) are each refined by the side refiners, \((c_h-1)\) jobs per level, to leave free exactly the days congruent to \(u\) modulo \(D\). Then on a day \(s\equiv u\pmod D\) whose residue modulo \(q\) is a marker residue, exactly one side is free, and in each of the other \(q-2\) residues modulo \(q\) both sides are free. One pair refiner of period \(qD\) in each of the latter \(q-2\) classes removes one of the two free slots.
\end{proof}

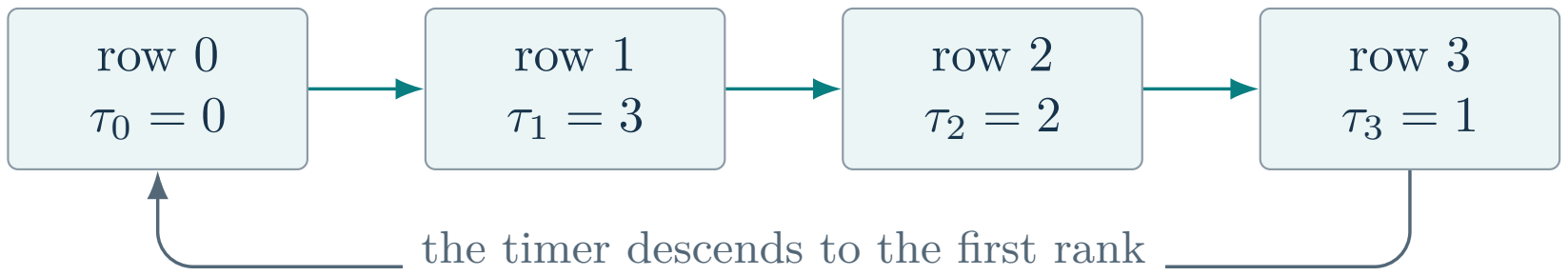
\Cref{fig:paired-example} illustrates the placement.

\subsection{Soundness: projecting the computation}\label{sec:projection}

\begin{figure}[t]
\centering
\begin{tikzpicture}[x=1cm,y=1cm,every node/.style={font=\small,text=ink}]
  \foreach \j/\rowres in {0/0,1/3,2/2,3/1} {
    \node[box,minimum width=2.08cm,fill=tealfig!8] (t\j) at (2.9*\j,0) {row \(\j\)\\ \(\tau_\j=\rowres\)};
  }
  \foreach \j/\k in {0/1,1/2,2/3} {\draw[flow,draw=tealfig] (t\j) -- (t\k);}
  \draw[flow,rounded corners=7pt] (t3.south) -- ++(0,-.67) -| (t0.south);
  \node[fill=white,font=\footnotesize,text=muted] at (4.35,-1.10) {the timer descends to the first rank};
\end{tikzpicture}
\smallskip

\small
\begin{tabular}{@{}llll@{}}
\toprule
Fine stream & Physical frame & Logical frame & Example for logical row 2 \\
\midrule
\(a=0\) & \(7z\) & \(z\) & logical frame \(6\) at physical frame \(42\) \\
\(a=1\) & \(1+7z\) & \(z+3\) & logical frame \(6\) at physical frame \(22\) \\
\bottomrule
\end{tabular}
\caption{Row ordering and stream alignment in a schematic example with \(M=4\), \(C\equiv3\pmod 4\), and \(B=7\). A slot at residue \(14\bmod 28\) and a different slot at residue \(22\bmod 28\) can serve row 2 in separate fine streams. One logical slot cannot use both residues, by \cref{lem:pair}.}\label{fig:row-streams}
\end{figure}

\begin{lemma}\label{lem:projection}
If \(\mathcal U\) is feasible, then \(\mathcal M\) is feasible.
\end{lemma}
\begin{proof}
Use the periodic schedule fixed above, with orientation \(\varepsilon\), row residues \(\tau_{\lambda_t}\equiv\tau_0+\varepsilon Ct\) (\cref{lem:timer}), and residues \(u_j\) (\cref{lem:pair}). Take a clean translate \(I'=\{s^*,\ldots,s^*+B(J+2M+2)-1\}\) for the computation jobs (\cref{lem:clean}); since the translation step \(MB\) is a multiple of \(B\), we may take \(s^*\equiv0\pmod B\). Inside \(I'\), every computation job occupies, on every day it is due, a slot of an application leaf outside all rigid footprints. If that slot lies in \(A_{j,0}\cup A_{j,1}\) for \(j=(t',x')\), then \(V_j(s)\ge1\), so \(s\equiv u_j\pmod D\).

\emph{Fine streams and logical days.} A computation job \(i\) is due only on days congruent to a fixed \emph{fine phase} \(a_i\) modulo \(B\), because \(B\) divides its period. For \(a\in\{0,\ldots,B-1\}\), let \(s_a\in\{0,\ldots,M-1\}\) be the unique residue with \(a\equiv\tau_0+\varepsilon Cs_a\pmod M\); it exists because \(C\) is a unit modulo \(M\). The day \(s=a+Bz\) receives the \emph{logical index} \(\kappa(s)=s_a+\varepsilon z\).

\emph{Rows are decoded correctly.} Suppose a computation job is at a slot of \(A_{j,0}\cup A_{j,1}\), \(j=(t',x')\), on day \(s=a+Bz\) of \(I'\). Then \(s\equiv u_j\equiv\tau_{\lambda_{t'}}\equiv\tau_0+\varepsilon Ct'\pmod M\). Since \(B\equiv C\pmod M\), also \(s\equiv\tau_0+\varepsilon Cs_a+Cz=\tau_0+\varepsilon C(s_a+\varepsilon z)\pmod M\). As \(C\) is a unit, \(t'\equiv\kappa(s)\pmod M\).

\emph{The decoded schedule.} For every computation job \(i\) and every day \(s\in I'\) on which it is due, put job \(i\) of \(\mathcal M\) on logical day \(\kappa(s)\) at hour \(x'\), where \(j=(t',x')\) is the logical slot it occupies. We check the conditions of \cref{def:masked}.
\begin{itemize}
\item \emph{Exact periods.} Job \(i\) is due on the days \(a_i+Bz\) with \(z\) in a fixed class modulo \(b_i\), so its logical days \(s_{a_i}+\varepsilon z\) form a class modulo \(b_i\), and each is used once.
\item \emph{Breaks.} The hour \(x'\) lies in \(S_{t'}\) with \(t'\equiv\kappa(s)\pmod M\).
\item \emph{Distinct hours.} Suppose that two jobs are decoded to the same logical day \(\kappa\) and hour \(x'\). They occupy slots of the same logical slot \(j=(\kappa\bmod M,x')\), on days \(s,s'\equiv u_j\pmod D\). Hence \(s\equiv s'\pmod B\), so they lie in the same fine stream \(a\), and \(\kappa(s)=\kappa(s')\) gives \(s=s'\). But \(V_j(s)=1\), so the two slots coincide, a contradiction.
\item \emph{Ascents.} Let \(\kappa\) and \(\kappa+b_i\) be consecutive logical visits of job \(i\), at logical hours \(x\) and \(x'\), and suppose \(x'-x\ge\delta_i+1\). A copy label of a slot with logical hour \(x\) lies in \(\{\pv x,\ldots,\pv x+E\}\). If \(\varepsilon=1\), the two visits come from physical days \(s\) and \(s+Bb_i\) in this order, and the physical hour equals the label by \cref{lem:layout}. The physical ascent is at least \(\pv(x'-x)-E\ge\pv\delta_i+\pv-E>\pv\delta_i+E\), since \(\pv=4M>2E\). If \(\varepsilon=-1\), the visit at \(\kappa+b_i\) comes from the physical day \(s-Bb_i\), which precedes the day \(s\) of the visit at \(\kappa\), and the physical hour of label \(c\) is \(w'+1-c\). The physical move from day \(s-Bb_i\) to day \(s\) is an ascent of at least \(\pv(x'-x)-E>\pv\delta_i+E\). Both cases contradict the ascent bound \(\pv\delta_i+E\).
\end{itemize}

\emph{Length.} The days of \(I'\) in fine stream \(a\) are \(a+Bz\) for \(z\) in an interval of \(J+2M+2\) consecutive integers, so their logical indices form an interval of the same length, shifted by \(s_a\in\{0,\ldots,M-1\}\). All streams therefore cover a common interval of at least \(J+M+3\) logical days, on which every job of \(\mathcal M\) is present on all its due days. Since \(J>b_*+Mw^n\), \cref{lem:frames}(a) yields a periodic feasible schedule of \(\mathcal M\).
\end{proof}

\Cref{fig:row-streams} illustrates how different fine streams serve the same logical row.

\subsection{Completeness}\label{sec:compiler-complete}

\begin{lemma}\label{lem:compiler-complete}
If \(\mathcal M\) is feasible, then \(\mathcal U\) is feasible.
\end{lemma}
\begin{proof}
Fix a feasible periodic schedule of \(\mathcal M\). We show how to place every job of \(\mathcal U\).

\emph{Hours and residues.} Put label \(\lam_i\) at hour \(i\). At every label, the free class of the labelling jobs is \(\ap0{\pi_\ell}\). Fix an injective assignment of residues modulo \(s_{\mathrm{lay}}\) to the layout groups and of residues modulo \(s_{\mathrm{row}}\) to the equality groups, the timer group, and the application group, avoiding \(0\); these exist by the bounds on the selectors. In the outer split, the layout reservation uses the residue \(1\) modulo \(s_0\) and the row reservation the residue \(2\). In an inner split, the leaf of group \(g\) uses the residue assigned to \(g\), and the guards take the remaining classes. In every chain, the free subclass at each level is the one congruent to \(0\) modulo the child prime. Set \(\tau_{\lambda_t}=Ct\bmod M\) and \(\tau_c=\tau_{\lambda_t}\) for every copy label \(c\) of row \(t\), and let the narrowing jobs cover the other classes modulo \(M\). With these choices, every aperture is determined by its residue modulo each prime factor of its modulus, and two leaves of the same group differ only in their hour-prime coordinates. Hence the leaves of every layout or equality group share a class modulo \(d_g\), and \cref{lem:tworank}(c) places the jobs of every such group. The leaves of a layout group lie at consecutive hours.

\emph{Timer.} All timer leaves agree in every coordinate except the hour prime and the residue modulo \(M\). Choose the phase \(\alpha_\theta\) of the timer job modulo \(C\) with \(\alpha_\theta\equiv0\pmod{\pi_{\lambda_t}}\) for every \(t\) and with the common residues of the timer leaves modulo \(s_0\), \(h_{\mathrm{row}}\), \(s_{\mathrm{row}}\), and \(H_{\mathrm{tim}}\). Fix an integer representative \(\alpha_\theta\) and let \(t_0\) satisfy \(Ct_0\equiv\alpha_\theta\pmod M\). On day \(\alpha_\theta+kC\), place \(\theta\) at the hour of \(\lambda_{t_0+k\bmod M}\). That day lies in the aperture of this timer leaf, its residue modulo \(M\) being \(C(t_0+k)\). From one visit to the next the hour rises by one, except from \(\lambda_{M-1}\) to \(\lambda_0\), which is a descent. The days used by \(\theta\) at \(\lambda_t\) form a class modulo \(MC\) inside the aperture of the timer leaf, and the timer refiners fill the rest of the aperture by mixed-radix refinement.

\emph{Application.} Choose a fine phase \(a\) modulo \(B\) with the common residues of the application leaves modulo \(s_0\), \(h_{\mathrm{row}}\), \(s_{\mathrm{row}}\), and \(H_{\mathrm{app}}\), with \(a\equiv0\pmod{\pi_c}\) for every copy label \(c\), and with \(a\equiv0\pmod m\). For every logical slot \(j=(t,x)\), let \(u_j\) be the class modulo \(D\) with \(u_j\equiv a\pmod B\) and \(u_j\equiv Ct\pmod M\). Both application leaves of \(j\) contain \(\ap{u_j}D\), and \cref{lem:pair-complete} places \(\mathcal F_j\).

\emph{Computation.} Let \(s_a\in\{0,\ldots,M-1\}\) satisfy \(a\equiv Cs_a\pmod M\), and map the logical day \(\kappa\) to the physical day \(\psi(\kappa)=a+B(\kappa-s_a)\). Since \(B\equiv C\pmod M\), we have \(\psi(\kappa)\equiv C\kappa\pmod M\), so \(\psi(\kappa)\equiv u_j\pmod D\) for every logical slot \(j\) of row \(\kappa\bmod M\). If job \(i\) of \(\mathcal M\) occupies hour \(x\) on logical day \(\kappa\), place the computation job \(i\) on day \(\psi(\kappa)\) on the unique free slot of the logical slot \((\kappa\bmod M,x)\). Distinct logical slots have disjoint copies, so no two jobs collide. Job \(i\) has period \(Bb_i\), because \(\psi\) multiplies differences by \(B\). Its physical hour is the label \(\pv x+2t+\sigma\) of the copy it uses, so a logical ascent of at most \(\delta_i\) becomes a physical ascent of at most \(\pv\delta_i+E\).

All structural jobs are placed as in their gadgets, and every group fills exactly its leaves. Hence the schedule is feasible.
\end{proof}

\subsection{Size of the instance}\label{sec:compiler-size}

\begin{lemma}\label{lem:compiler-size}
The instance \(\mathcal U\) has \(O\bigl((Mw)^4\log^2(Mw)+n(Mw)^2\log w\bigr)\) jobs, every period has \(O(Mw\log(Mw)+n\log w+\log n)\) bits, and \(\mathcal U\) can be computed in time polynomial in \(M\), \(w\), and the input length.
\end{lemma}
\begin{proof}
There are \(w'=O(Mw)\) labels, \(O(Mw)\) groups, and at most \(Mw\) logical slots, so \(O(Mw)\) primes are needed. Taking the smallest primes above \(\max(M,11)\), all of them are \(O(Mw\log(Mw))\) and can be found by a sieve. Since \(\log\Theta=O(n\log w+\log(nM))\), every exponent \(a_g\) is polynomial, and \(m\le M\) can be factored by trial division. The labelling jobs number \(\sum_\ell(\pi_\ell-1)=O((Mw)^2\log(Mw))\), as do the guards and the jobs of the two-hour tests. The chains contribute \(O(N_{\mathrm{leaf}}\max_gh_ga_g)=O((Mw)^2(n\log w+\log(nM)))\) jobs, and the timer refiners \(O(M^2\cdot Mw\log(Mw))\). For each logical slot and side, the side refiners number \((q_j-1)\sum_h(c_h-1)\), where the \(c_h\) range over the prime factors of \(m\prod_{c\ne c_s}\pi_c\), which gives \(O(Mw\log(Mw)\cdot(Mw)^2\log(Mw))\) per slot and \(O((Mw)^4\log^2(Mw))\) in total. Every period is a product of \(O(Mw)\) of the primes, of one or two group powers, of \(M\), \(m\), and of at most \(K\), which gives the bound on the bit length.
\end{proof}

\begin{proof}[Proof of \cref{thm:compiler}]
The instance \(\mathcal U\) is saturated by \cref{lem:saturation} and has polynomial size by \cref{lem:compiler-size}. It is feasible exactly when \(\mathcal M\) is, by \cref{lem:projection,lem:compiler-complete}.
\end{proof}

\section{From Daily Scheduling to Pinwheel Scheduling}\label{sec:dyadic}

\begin{theorem}\label{thm:dyadic}
There is a polynomial-time reduction from saturated Daily Scheduling, with an explicit list of jobs, to Pinwheel Scheduling with an explicit list of jobs and binary deadlines. If some ascent bound is positive, the output density is strictly between \(1-n\Delta/T^2\) and \(1\), where \(T\) is the day length defined below.
\end{theorem}

Let the input have width \(w\) and jobs \((b_i,\delta_i)_{i\in[n]}\) with \(\sum_i1/b_i=w\). Since every \(b_i\ge1\), saturation gives \(w\le n\). Define
\[
L=\lcm_ib_i,\qquad b_*=\max_ib_i,\qquad \Delta=\max_i\delta_i,\qquad \Omega=Lw^n+6L+10,
\]
\[
\mathcal E=\Omega(\Delta+1)\bigl(n(b_*+1)+w\bigr).
\]
The parameter \(\Omega\) is the number of days we will recover, and \(\mathcal E\) bounds the total deviation from exact days in a window of \(\Omega\) days. Choose the least \(k\ge1\) such that \(T=w2^k\) satisfies \(T>\Delta\) and \(T>n(2\mathcal E+3)\); any larger \(k\) with polynomially many bits also works.

\paragraph{The instance.}
The output has a \emph{payload job} for every input job, with deadline
\begin{equation}\label{eq:ordinary-deadlines}
d_i=Tb_i+\delta_i,
\end{equation}
and \(w\) \emph{clock jobs of level \(h\)} with deadline \(2^hw\), for every \(h=1,\ldots,k\). The bit length of \(L\) is at most the total bit length of the \(b_i\), and \(\log w^n=n\log w\), so \(\log T\) is polynomial, and the \(wk\) clock jobs can be listed explicitly. The clock jobs have density \(\sum_{h=1}^kw/(2^hw)=1-2^{-k}=1-w/T\).

\subsection{Completeness}

\begin{lemma}\label{lem:dyadic-complete}
If the Daily Scheduling instance is feasible, the pinwheel instance is feasible.
\end{lemma}
\begin{proof}
Partition the time axis into \emph{blocks} of \(w\) consecutive slots, indexed by \(\Z\). Every nonzero integer \(z\) has a unique \(h\ge1\) with \(z\equiv2^{h-1}\pmod{2^h}\). For \(h\le k\), place the \(w\) clock jobs of level \(h\) in the blocks with \(z\equiv2^{h-1}\pmod{2^h}\), always in the same order; each then has gap exactly \(2^hw\). The blocks left free are those with \(z\equiv0\pmod{2^k}\), one every \(T\) slots. Let block \(2^kt\) represent day \(t\), and let its \(y\)-th slot represent hour \(y\). Place a feasible schedule for that instance there. If a job uses hour \(y\) on day \(t\) and hour \(y'\) on day \(t+b_i\), its gap is \(Tb_i+y'-y\le Tb_i+\delta_i=d_i\).
\end{proof}

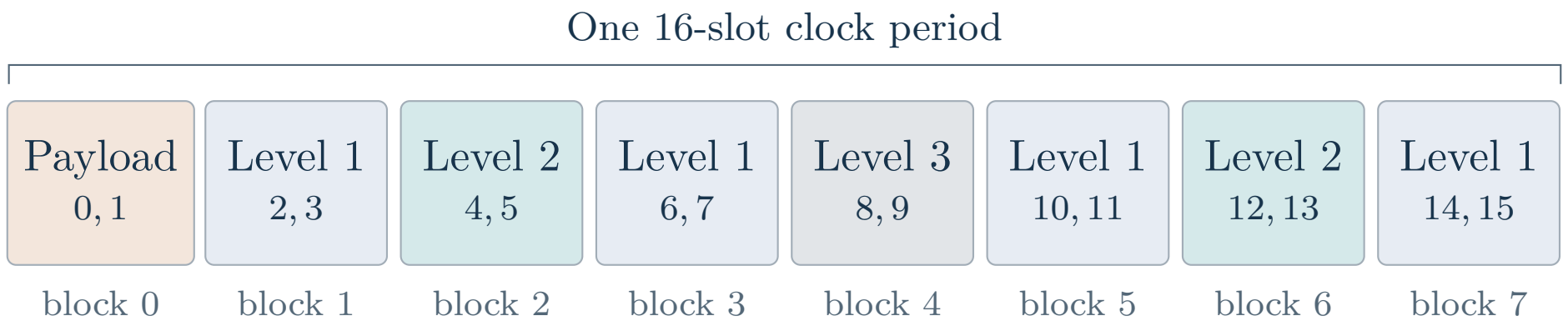
\begin{figure}[t]
\centering
\begin{tikzpicture}[x=1.55cm,y=.72cm,every node/.style={font=\small}]
  \foreach \block/\col/\name in {0/ochre!15/Payload,1/bluefig!12/Level 1,2/tealfig!17/Level 2,3/bluefig!12/Level 1,4/muted!17/Level 3,5/bluefig!12/Level 1,6/tealfig!17/Level 2,7/bluefig!12/Level 1}{
    \pgfmathtruncatemacro{\lefttime}{2*\block}
    \pgfmathtruncatemacro{\righttime}{2*\block+1}
    \node[draw=ink!40,fill=\col,rounded corners=2pt,minimum width=1.44cm,minimum height=1.3cm,align=center,text=ink]
        at (\block,0) {\name\\[-1pt]\scriptsize \(\lefttime,\righttime\)};
    \node[below,font=\scriptsize,text=muted] at (\block,-1) {block \(\block\)};
  }
  \draw[muted] (-.47,1.09) -- (-.47,1.30) -- (7.47,1.30) -- (7.47,1.09);
  \node[above,text=ink,font=\footnotesize] at (3.5,1.30) {One 16-slot clock period};
\end{tikzpicture}
\caption{The canonical dyadic placement with \(w=2\) and \(T=16\). The two jobs of a level always take the same order in their blocks. Levels 1, 2, and 3 have deadlines \(4,8,16\), respectively. A payload of period two that starts at rank 1 and next uses rank 2 runs at slots \(0\) and \(33\), with gap \(2T+1\).}\label{fig:dyadic-example}
\end{figure}

\subsection{Soundness}

Fix a feasible pinwheel schedule. By \cref{prop:space} we may assume that it is periodic with some period \(\Pi\). Call a slot a \emph{payload slot} if it is not occupied by a clock job; idle slots count as payload slots.

\begin{lemma}\label{lem:spacing}
Let \(\cdots<x_{-1}<x_0<x_1<\cdots\) be the payload slots. Then \(x_{j+w}-x_j\ge T\) for every \(j\). In particular, every \(T\) consecutive slots contain at most \(w\) payload occurrences.
\end{lemma}
\begin{proof}
Delete the clock jobs level by level, in increasing order of level, keeping the time coordinates of the remaining slots. Let \(\xi^{(h)}\) denote the increasing sequence of slots that remain after levels \(1,\ldots,h\) have been deleted, so \(\xi^{(0)}\) is all of \(\Z\). We show by induction that \(\xi^{(h)}_{j+w}-\xi^{(h)}_j\ge2^hw\) for all \(j\). For \(h=0\) this is an equality. Assume it for \(h-1\). Consider consecutive occurrences of a level-\(h\) clock job at positions \(\xi^{(h-1)}_a<\xi^{(h-1)}_b\). If \(b-a\ge2w+1\), then
\[
\xi^{(h-1)}_b-\xi^{(h-1)}_a\ \ge\ \bigl(\xi^{(h-1)}_{a+2w}-\xi^{(h-1)}_{a+w}\bigr)+\bigl(\xi^{(h-1)}_{a+w}-\xi^{(h-1)}_a\bigr)+1\ \ge\ 2^hw+1,
\]
which exceeds the deadline \(2^hw\). Hence every \(2w\) consecutive entries of \(\xi^{(h-1)}\) contain an occurrence of each of the \(w\) clock jobs of level \(h\), and at most \(w\) entries that survive into \(\xi^{(h)}\). Now let \(\xi^{(h)}_j=\xi^{(h-1)}_a\) and \(\xi^{(h)}_{j+w}=\xi^{(h-1)}_b\). The \(w+1\) surviving entries \(\xi^{(h)}_j,\ldots,\xi^{(h)}_{j+w}\) lie among \(\xi^{(h-1)}_a,\ldots,\xi^{(h-1)}_b\), so \(b-a+1>2w\), that is, \(b-a\ge2w\). By the induction hypothesis applied twice, \(\xi^{(h)}_{j+w}-\xi^{(h)}_j\ge2\cdot2^{h-1}w=2^hw\). For \(h=k\) we obtain the claim.
\end{proof}

Let \(f_i\) be the frequency of payload job \(i\), that is, its number of occurrences per period divided by \(\Pi\).

\begin{lemma}\label{lem:frequency}
We have \(f_i\ge1/d_i\) for every \(i\), \(\sum_if_i\le w/T\), and
\begin{equation}\label{eq:frequency}
\sum_i\Bigl(f_i-\frac1{d_i}\Bigr)\ \le\ \frac wT-\sum_i\frac1{Tb_i+\delta_i}\ =\ \sum_i\frac{\delta_i}{Tb_i(Tb_i+\delta_i)}\ \le\ \frac{n\Delta}{T^2}.
\end{equation}
\end{lemma}
\begin{proof}
All gaps of job \(i\) are at most \(d_i\), so \(f_i\ge1/d_i\). By \cref{lem:spacing}, a period contains at most \(w\Pi/T\) payload occurrences, so \(\sum_if_i\le w/T\). The equality uses saturation, \(w/T=\sum_i1/(Tb_i)\).
\end{proof}

At each occurrence of payload job \(i\), let \(g\) be its \emph{forward gap}, the distance to its next occurrence. Its \emph{deviation} is \(|g-Tb_i|\). Since \(g\le d_i\),
\begin{equation}\label{eq:deviation}
|g-Tb_i|\ \le\ d_i-g+\delta_i ,
\end{equation}
because if \(g\ge Tb_i\) then \(g-Tb_i\le\delta_i\), and if \(g<Tb_i\) then \(Tb_i-g=d_i-g-\delta_i\).

\begin{lemma}\label{lem:window}
There is a window of \(\Omega T\) consecutive slots such that the total deviation of the payload occurrences in it is at most \(\mathcal E\).
\end{lemma}
\begin{proof}
Over one period, the gaps of job \(i\) add up to \(\Pi\), so \(\sum(d_i-g)=\Pi(f_id_i-1)\) over its occurrences, and \(\sum\delta_i=\Pi f_i\delta_i\). Choose the window start uniformly at random among the \(\Pi\) slots of a period. By linearity of expectation and \cref{eq:deviation}, the expected total deviation of the occurrences in the window is at most
\[
\Omega T\sum_i\bigl(f_id_i-1+f_i\delta_i\bigr)\ \le\ \Omega T\Bigl(\max_id_i\sum_i\Bigl(f_i-\frac1{d_i}\Bigr)+\Delta\sum_if_i\Bigr)\ \le\ \Omega\Delta\bigl(n(b_*+1)+w\bigr)\ \le\ \mathcal E,
\]
using \(f_i-1/d_i\ge0\), \(\max_id_i\le T(b_*+1)\) because \(T>\Delta\), and \cref{lem:frequency}. Some window attains at most the expectation.
\end{proof}

\begin{lemma}\label{lem:cut}
Inside the window of \cref{lem:window}, there is an integer \(\gamma\) such that, for the days \(\{\gamma+tT,\ldots,\gamma+(t+1)T-1\}\), consecutive occurrences of every payload job \(i\) lie exactly \(b_i\) days apart.
\end{lemma}
\begin{proof}
For each job, list its occurrences in the window as \(x^{(0)}<x^{(1)}<\cdots\). Then \(x^{(\iota)}=x^{(0)}+\iota Tb_i+D_\iota\), where the cumulative deviation \(D_\iota\) is a sum of deviations of distinct occurrences, so \(|D_\iota|\le\mathcal E\). Hence the residues of the occurrences of job \(i\) modulo \(T\) lie in the cyclic interval of radius \(\mathcal E\) around \(x^{(0)}\bmod T\). These \(n\) intervals contain at most \(n(2\mathcal E+1)<T\) residues, so some residue \(\gamma\bmod T\) lies in none of them. For this cut, the integers \(x^{(0)}+D\) with \(|D|\le\mathcal E\) all lie strictly inside one day, so \(x^{(\iota)}\) lies on the day that contains \(x^{(0)}\), shifted by \(\iota b_i\) days.
\end{proof}

Let the window cover the days \(t=t_1,\ldots,t_2\) of this cut, where the two extreme days may be partial. Discard them and \(L+2\) further complete days at each end, and call the rest the \emph{retained days}. Their number is at least \(\Omega-2-2(L+2)\ge Lw^n+4L+4\). Every occurrence on a retained day has its forward and backward gaps inside the window, because a gap is less than \(T(b_*+1)\le T(L+1)\). So each payload job \(i\) occurs exactly once on each retained day of one class modulo \(b_i\), and on no other retained day.

\begin{lemma}[Full days]\label{lem:full}
Every retained day contains exactly \(w\) payload occurrences.
\end{lemma}
\begin{proof}
A day has \(T\) slots, so it contains at most \(w\) payload occurrences by \cref{lem:spacing}. The number \(N(t)\) of jobs whose class contains \(t\) is periodic with period \(L\) and has average \(\sum_i1/b_i=w\). The retained days contain at least \(L\) consecutive days, so \(N(t)=w\) on a complete period and hence on all retained days.
\end{proof}

\begin{lemma}[Hours]\label{lem:ranks}
Assign hours \(1,\ldots,w\) to the payload occurrences of each retained day in increasing order of time. Then a job whose hour rises by \(a\ge0\) between two consecutive retained visits satisfies \(a\le\delta_i\).
\end{lemma}
\begin{proof}
Let \(\eta_{t,y}\) be the offset, from the start \(\gamma+tT\) of day \(t\), of the payload occurrence assigned hour \(y\). A retained day has \(T\) slots and \(w\) payload occurrences, so by \cref{lem:spacing} it contains no other payload slot, in particular no idle slot. Hence the occurrence at hour \(y\) on day \(t+1\) is the \(w\)-th payload slot after the occurrence at hour \(y\) on day \(t\), and \cref{lem:spacing} gives \((\gamma+(t+1)T+\eta_{t+1,y})-(\gamma+tT+\eta_{t,y})\ge T\). So \(\eta_{t+1,y}\ge\eta_{t,y}\), and by induction \(\eta_{t',y}\ge\eta_{t,y}\) for retained days \(t'\ge t\). Within a day the offsets are distinct integers, so \(\eta_{t,y+a}\ge\eta_{t,y}+a\). If job \(i\) is at hour \(y\) on day \(t\) and at hour \(y+a\) on day \(t+b_i\), its gap is
\[
Tb_i+\eta_{t+b_i,y+a}-\eta_{t,y}\ \ge\ Tb_i+\eta_{t,y+a}-\eta_{t,y}\ \ge\ Tb_i+a .
\]
Its deadline \(Tb_i+\delta_i\) gives \(a\le\delta_i\).
\end{proof}

\begin{lemma}\label{lem:dyadic-sound}
If the pinwheel instance is feasible, the Daily Scheduling instance is feasible.
\end{lemma}
\begin{proof}
The retained days, with the hours of \cref{lem:ranks}, form a schedule that satisfies the conditions of \cref{def:ref} on more than \(b_*+Lw^n\) consecutive days: every job has an exact phase, the hours on a day are distinct, and every ascent is legal. By \cref{lem:frames}(a), the Daily Scheduling instance has a periodic feasible schedule.
\end{proof}

\begin{proof}[Proof of \cref{thm:dyadic}]
Completeness and soundness are \cref{lem:dyadic-complete,lem:dyadic-sound}, and the size bounds were shown above. The density is \(1-w/T+\sum_i1/(Tb_i+\delta_i)\), so by \cref{eq:frequency},
\[
1-\rho=\sum_i\frac{\delta_i}{Tb_i(Tb_i+\delta_i)}\in\Bigl(0,\ \frac{n\Delta}{T^2}\Bigr]
\]
whenever some \(\delta_i\) is positive.
\end{proof}

\section{The main theorem and its consequences}\label{sec:consequences}

\begin{proof}[Proof of \cref{thm:main}]
Membership in \(\PSPACE\) is due to~\cref{prop:space}. For hardness, start from Periodic 3-SAT, which is \(\PSPACE\)-complete by \cref{lem:psat}. \Cref{thm:masked} produces a saturated instance with breaks, of polynomial width and calendar period and with job periods dividing \(M\). \Cref{thm:compiler} produces an equivalent saturated instance without breaks, specified by an explicit list of polynomially many jobs, and \cref{thm:dyadic} produces an equivalent pinwheel instance with an explicit list of polynomially many jobs and deadlines of polynomial bit length. Each reduction runs in polynomial time and preserves feasibility in both directions.
\end{proof}

\begin{proof}[Proof of \cref{thm:intro-near}]
Membership follows from \cref{prop:space}, because the density condition can be checked in polynomial time with exact rational arithmetic. For hardness, the instance without breaks produced by \cref{thm:compiler} contains the timer job, whose ascent bound is \(1\), so \(\Delta\ge1\). By \cref{thm:dyadic}, the output density satisfies \(1-n\Delta/T^2<\rho<1\). Increasing \(k\) by \(O(\log(n\Delta/\epsilon))\) makes \(n\Delta/T^2<\epsilon\); this adds polynomially many clock jobs and bits, and the proof of \cref{thm:dyadic} applies to any \(k\) above its threshold.
\end{proof}

\begin{proof}[Proof of \cref{cor:intro-consequences}]
Every problem in \(\NP\) or \(\coNP\) is in \(\PSPACE\) and therefore reduces to Pinwheel Scheduling. If Pinwheel Scheduling were in \(\NP\), then every problem in \(\PSPACE\) would be in \(\NP\), so \(\NP=\PSPACE\). The statement for \(\coNP\) is symmetric.
\end{proof}

\begin{corollary}\label{cor:compact}
Pinwheel Scheduling is \(\PSPACE\)-hard when equal deadlines are grouped and their multiplicities are written in binary.
\end{corollary}
\begin{proof}
Grouping the explicit list produced by \cref{thm:main} takes polynomial time and describes the same instance.
\end{proof}

We do not know whether the compact version lies in \(\PSPACE\), because its state vectors (cf.~\cite{holte1989,kanellopoulos2026icalp}) can have exponentially many entries.

\paragraph{Density one is different.}
The following observation, essentially due to \citet{holte1989}, explains why our instances must have density below one.

\begin{proposition}\label{prop:density-one}
Pinwheel Scheduling restricted to instances of density exactly one is in \(\NP\).
\end{proposition}
\begin{proof}
By \cref{prop:space}, a feasible instance has a periodic schedule. Its frequencies satisfy \(f_i\ge1/d_i\) and \(\sum_if_i\le1=\sum_i1/d_i\), so \(f_i=1/d_i\) for every \(i\). Since every gap of job \(i\) is at most \(d_i\) and the gaps average \(d_i\), every gap equals \(d_i\). Hence a certificate consists of one phase per job modulo its deadline, and it can be verified by checking every pair of jobs with \cref{fact:crt}; full coverage then follows from density one.
\end{proof}

\section*{Acknowledgement of AI use}
The authors have been working on the complexity of Pinwheel Scheduling for the past two years. The key breakthrough behind the reduction presented here came during a conversation with OpenAI's ChatGPT 6.2~Sol on November~18, 2026. The authors subsequently verified the validity of the result and, with the assistance of Anthropic's Claude Opus~5.5, worked on simplifying the exposition of the paper. The authors take full responsibility for the contents of this paper.

\bibliographystyle{plainnat}
\bibliography{references}

\newpage
\appendix
\section{Notation and parameters}\label{app:notation}

\begin{table}[h]
\centering
\small
\begin{tabularx}{\linewidth}{@{}l X@{}}
\toprule
Symbol & Meaning \\
\midrule
\(\ap am\) & the class \(\{t\in\Z:t\equiv a\pmod m\}\) \\
\(b_i,\ \delta_i\) & period (in days) and ascent bound of job \(i\) \\
\(w\) & instance width (\(w'\) for the compiler output) \\
\(M,\ S_t\) & calendar period and rows of an instance with breaks \\
\(N,\ m,\ v\) & variables per copy, clauses, strips \(v=2N\) (\cref{sec:masked}) \\
\(o_\sigma,\ R_\sigma\) & low hour and region of strip \(\sigma\) (\cref{sec:masked}) \\
\(W,\ G\) & width of the middle band and guard offset, \(W=12v+2\), \(G=W+4\) \\
\(\mathsf A,\ \mathsf B\) & the two banks \\
\(\pv,\ E\) & pitch \(4M\) and span \(2M-1\) of the copies of a logical slot (\cref{sec:compiler}) \\
\(\lam,\ \phi\) & label sequence and the label-to-hour bijection \\
\(\pi_\ell,\ s_0,\ s_{\mathrm{lay}},\ s_{\mathrm{row}}\) & hour primes and selectors \\
\(h_{\mathrm{lay}},\ h_{\mathrm{row}},\ h_g,\ H_g\) & branch primes, group primes, group powers \(H_g=h_g^{a_g}\) \\
\(q_j\) & pair prime of logical slot \(j\) \\
\(\tau_\ell\) & row residue of label \(\ell\) modulo \(M\) \\
\(C,\ B,\ D\) & period of the timer job, unit of the computation periods, \(D=MB\) \\
\(K,\ J,\ \Theta\) & \(2310\), the recovered length \(Mw^n+2M+10\), the bound on group powers \\
\(L,\ \Omega,\ \mathcal E,\ T\) & \(\lcm_ib_i\), recovered days, deviation budget, day length (\cref{sec:dyadic}) \\
\bottomrule
\end{tabularx}
\caption{Notation.}\label{tbl:notation}
\end{table}

\begin{table}[h]
\centering
\small
\begin{tabularx}{\linewidth}{@{}l l l X@{}}
\toprule
Family & Number & Period & Ascent bound and role \\
\midrule
Labelling of \(\ell\) & \(\pi_\ell-1\) & \(\pi_\ell\) & rigid; pins label \(\ell\) to an hour \\
Guards of a split & \(s-k\) & \(sQ\) & rigid; unused selector classes \\
Chain of a child & \((h-1)a\) & \(sQh^i\), \(i\in[a]\) & rigid; leaves one class modulo\hspace{-0.25pt} \(sQh^a\) \\
Row narrowing at \(\ell\) & \(M-1\) & \(MQ_\ell\) & rigid; fixes the row residue \(\tau_\ell\) \\
Test refiners of \(g\) & \(\pi_\ell+\pi_{\ell'}-2\) & \(d_g\) & rigid; leave two slots per \(d_g\) days \\
Witnesses of \(g\) & \(5\) & \(d_gc_v\) & \(1\) (layout) or \(w'-1\) (equality) \\
Test fillers of \(g\) & \(4002\) & \(d_gK\) & rigid \\
Timer refiners & \(\sum_t\sum_{t'\ne t}(\pi_{\lambda_{t'}}-1)\) & divisors of \(MC\) & rigid \\
Timer job & \(1\) & \(C\) & \(1\) \\
Markers of \(j\) & \(2\) & \(q_jP_s\) & rigid; identify the pair \\
Side refiners of \(j\) & \((q_j-1)\sum_h(c_h-1)\) per side & \(q_jP_sc_1\cdots c_h\) & rigid \\
Pair refiners of \(j\) & \(q_j-2\) & \(q_jD\) & rigid \\
Computation job \(i\) & \(1\) & \(Bb_i\) & \(\pv\delta_i+E\) \\
\bottomrule
\end{tabularx}
\caption{The job families of the compiled instance \(\mathcal U\) (\cref{sec:compiler}).}\label{tbl:families}
\end{table}

\end{document}